\documentclass[article,aps,nofootinbib,twocolumn,superscriptaddress]{revtex4-1}
\input epsf
\usepackage{graphics}
\usepackage{enumitem}
\usepackage{amsmath}
\usepackage{amssymb}
\usepackage{bm}
\usepackage{braket}
\usepackage{booktabs}
\usepackage{subfigure}
\usepackage{subfloat}
\usepackage{float}
\usepackage[utf8]{inputenc}
\usepackage[T1]{fontenc}
\usepackage{currvita}

\usepackage{hyperref}

\hypersetup{
    colorlinks=true,
    linkcolor=red,
    citecolor=blue,
}

\usepackage{color}
\usepackage{dcolumn}
\usepackage{hyphenat}

\def\be{\begin{equation}}
\def\ee{\end{equation}}
\def\ba{\begin{eqnarray}}
\def\ea{\end{eqnarray}}

\def\ie{{\frenchspacing\it i.e.}}
\def\eg{{\frenchspacing\it e.g.}}
\def\etc{{\frenchspacing\it etc.}}

\newcommand{\gb}[1]{{\bf \color{blue}{{#1}}}}

\usepackage{graphicx}

\begin{document}

\title{Generalized Brans-Dicke theories in light of evolving dark energy}

\author{Alex Zucca}  \affiliation{Department of Physics, Simon Fraser University, Burnaby, BC, V5A 1S6, Canada}
\author{Levon Pogosian} \affiliation{Department of Physics, Simon Fraser University, Burnaby, BC, V5A 1S6, Canada} \affiliation{Institute of Cosmology and Gravitation, University of Portsmouth, Portsmouth, PO1 3FX, UK}
\author{Alessandra Silvestri} \affiliation{Institute Lorentz, Leiden University, PO Box 9506, Leiden 2300 RA, The Netherlands}
\author{Yuting Wang} \affiliation{National Astronomy Observatories, Chinese Academy of Sciences, Beijing, 100101, P.~R.~China} 
\author{Gong-Bo Zhao} \affiliation{National Astronomy Observatories, Chinese Academy of Sciences, Beijing, 100101, P.~R.~China} 
\affiliation{School of Astronomy and Space Science, University of Chinese Academy of Sciences, Beijing 100049, P.~R.~China}
\affiliation{Institute of Cosmology and Gravitation, University of Portsmouth, Portsmouth, PO1 3FX, UK}

\begin{abstract}
The expansion history of the Universe reconstructed from a combination of recent data indicates a preference for a changing Dark Energy (DE) density. Moreover, the DE density appears to be increasing with cosmic time, with its equation of state being below $-1$ on average, and possibly crossing the so-called phantom divide. Scalar-tensor theories, in which the scalar field mediates a force between matter particles, offer a natural framework in which the effective DE equation of state can be less than $-1$ and cross the phantom barrier. We consider the generalized Brans-Dicke (GBD) class of scalar-tensor theories and reconstruct their Lagrangian given the effective DE density extracted from recent data. Then, given the reconstructed Lagrangian, we solve for the linear perturbations and investigate the characteristic signatures of these reconstructed GBD in the cosmological observables, such as the cosmic microwave background (CMB) anisotropy, the galaxy number counts, and their cross-correlations. In particular, we demonstrate that the Integrated Sachs-Wolfe (ISW) effect probed by the cross-correlation of CMB with the matter distribution can rule out scalar-tensor theories as the explanation of the observed DE dynamics independently from the laboratory and solar system fifth force constraints.

\end{abstract}

\maketitle

\section{Introduction}
The observed accelerated expansion of the Universe has been puzzling cosmologists since its discovery two decades ago \citep{Perlmutter:1998np, Riess:1998cb}. Within the context of General Relativity (GR), it implies the existence of an energy-momentum component with a negative equation of state (EOS), referred to as Dark Energy (DE). The standard cosmological model, $\Lambda$CDM, in which DE is the constant energy of the vacuum, provides a good fit to a plethora of cosmological observations such as the cosmic microwave background (CMB) anisotropies  \citep{WMAP, Ade:2015xua}, baryon acoustic oscillations (BAO) \citep{Percival:2002gq,6df,Alam:2016hwk}, type Ia supernovae \citep{Conley:2011ku, Suzuki:2011hu}, galaxy clustering \citep{SDSS} and galaxy lensing \citep{Heymans:2012gg, Hildebrandt:2016iqg}. However, $\Lambda$CDM is not fully satisfactory from the theoretical perspective, as the observed value of the vacuum energy requires an extreme fine tuning of the cosmological constant $\Lambda$ in the context of the present understanding of particle interactions \citep{Burgess:2017ytm}. Also, with the data becoming more accurate, several ``tensions'' between different datasets have begun to arise when interpreting observations within the $\Lambda$CDM model \cite{Ade:2015xua,Riess:2016jrr, Riess:2019cxk, Abbott:2017wau,Wong:2019kwg}. Although these tensions might just be due to unaccounted systematic effects or rare statistical fluctuations \citep{Scott:2018adl}, they generated significant interest in possible extensions of $\Lambda$CDM capable of relieving the tensions \cite{Dvorkin:2014lea,DiValentino:2016hlg,DiValentino:2017oaw,DiValentino:2017rcr,Adhikari:2019fvb,Lin:2019qug,Peirone:2019aua}, including the possibility of the DE density evolving with time \cite{Zhao:2017cud,Wang:2018fng,Yang:2018uae,DiValentino:2019exe}. 

Using a combination of available observations, non-parametric reconstructions of the DE dynamics were performed in \citep{Zhao:2017cud,Wang:2018fng}. Interestingly, they show a preference for an \emph{increasing} effective DE density, {\it i.e.} one with an EOS, $\smash{w_{\rm DE}^{\rm eff}<-1}$. The reconstruction also shows crossing of the so-called phantom divide \cite{Caldwell:1999ew, Carroll:2003st, Vikman:2004dc} of $\smash{w_{\rm DE}^{\rm eff}=-1}$. Such dynamics cannot be explained by a minimally coupled quintessence field DE, but could be realized in scalar-tensor extensions of GR where the additional scalar field $\phi$ mediates a force between particles \cite{Carroll:2003st,Das:2005yj,Amendola:2007nt}. In fact, scalar-tensor theories possess enough freedom to reproduce any expansion history.

The aim of this paper is to investigate scalar-tensor theories of the generalized Brans-Dicke (GBD) type capable of realizing the expansion histories reconstructed in \citep{Zhao:2017cud,Wang:2018fng}. Using the observed $H(a)$ as input, and making certain assumptions about the scalar field coupling function, we systematically scan the parameter space to reconstruct the GBD Lagrangians consistent with that $H(a)$. We then solve for the cosmological perturbations and calculate predictions for the CMB and galaxy power spectra and other observables to isolate the theories that are in agreement with current data. 

Late time deviations from $\Lambda$CDM are mainly encoded in the CMB temperature through the integrated Sachs-Wolfe (ISW) effect. Although too small to be detected from the CMB temperature auto-correlation, the ISW contribution can be probed by cross correlating the CMB temperature maps with the foreground galaxies number counts \citep{1996PhRvL..76..575C, Afshordi:2004kz, Giannantonio:2008zi}, which can be a useful probe for DE \cite{Pogosian:2004wa,Pogosian:2005ez}. In $\Lambda$CDM, the accelerating expansion results in decaying gravitational potentials, yielding a strictly positive ISW effect. In scalar-tensor theories however, the ISW effect can have a positive or negative sign depending on whether the enhanced clustering due to the fifth force, which yields a negative ISW, dominates over the effect of the accelerating expansion \cite{Song:2006ej,Renk:2016olm}. We find that most of the GBD theories reconstructed in this work predict CMB-matter cross-correlations that are significantly different from those in $\Lambda$CDM and, therefore, can be ruled out or confirmed with the next generation galaxy surveys, such as DESI \cite{DESI}, LSST \cite{Ivezic:2008fe, 2009arXiv0912.0201L} and Euclid \cite{Refregier:2008js, Cimatti:2008kc, 2009arXiv0912.0914L}. 

The paper is structured as follows. In Sec.~\ref{sect:DERecon} we discuss the DE reconstruction results from \citep{Wang:2018fng} and introduce expansion models used in this work, along with a brief review of GBD theories. In Sec.~\ref{sect:GBDtheories} we introduce the GBD theories and describe two different ways in which we have reconstructed their Lagrangians from a given expansion history. Then, in Sec.~\ref{sec:observables}, we evaluate the key cosmological observables predicted by the viable GBD theories that includes the CMB-galaxy cross-correlation spectra at three representative redshifts, showing how the latter can help to rule out or confirm GBD as an alternative to $\Lambda$CDM. We summarize our findings in Sec.~\ref{sec:summary}.

\section{The reconstructed dark energy density}
\label{sect:DERecon}

\begin{figure*}[tbph]
\centering
\includegraphics[width = .9\textwidth]{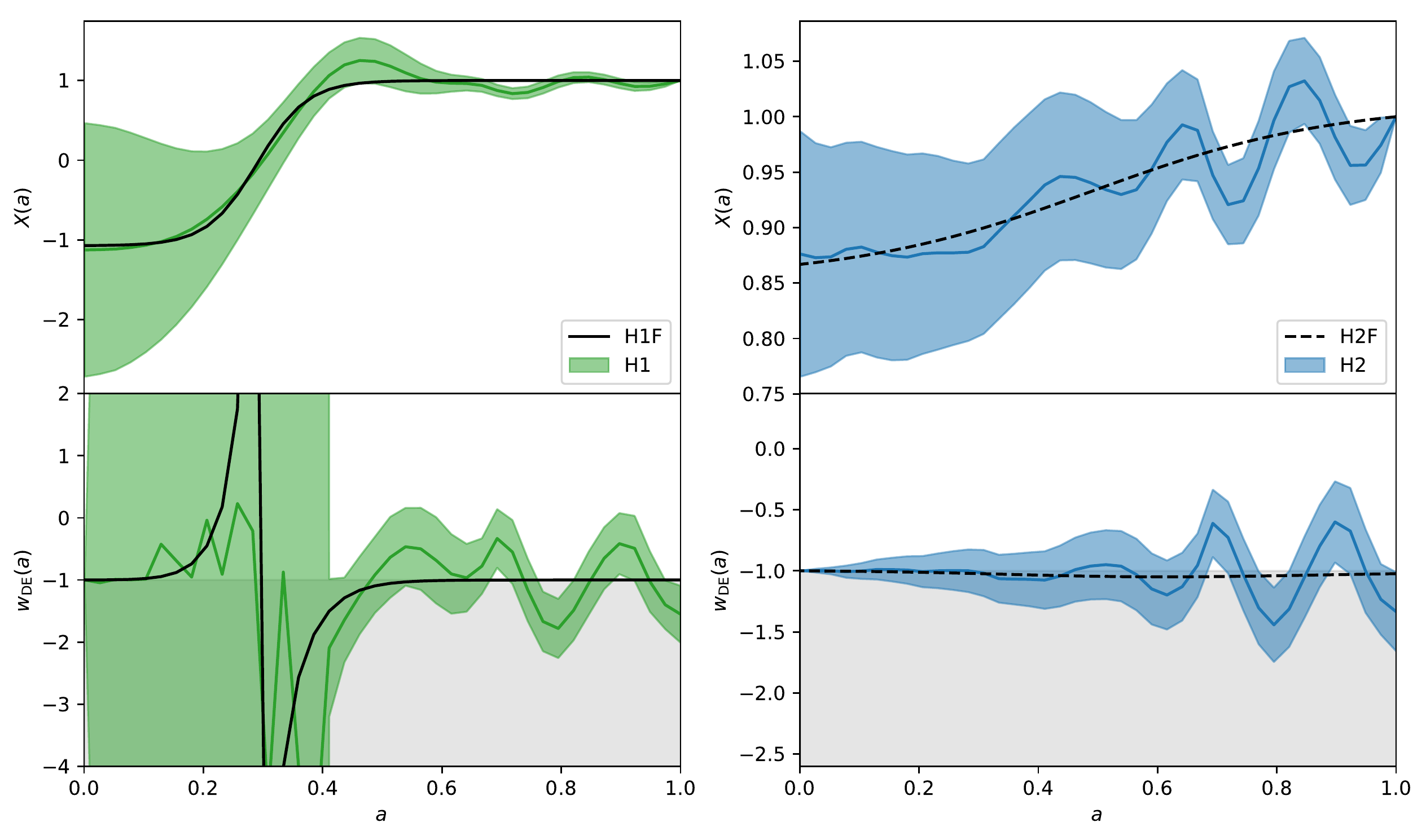}
\caption{\label{fig:xDE_wDE_rec} The upper panels show the reconstructed normalized effective DE density, $X(a)$, obtained using the standard prior (H1, left) and the evidence-weighted method (H2, right). Also shown are the corresponding hyperbolic tangent fits H1F (solid line) and H2F (dashed line). The lower panels show the corresponding effective DE equation of state.}
\end{figure*}

A Bayesian, non-parametric reconstruction of the time-evolution of the DE density was performed in \citep{Wang:2018fng} using the correlated prior method introduced in \citep{2012JCAP...02..048C, Crittenden:2005wj}. The effective DE energy density is modelled through the dimensionless function $X(a)$ that enters the Friedmann equation via
\begin{equation}
\label{eqn:FriedmannPhenomen}
H^2 = H_0^2 \left[ E_m(a) +\Omega_{\Lambda} X(a) \right].
\end{equation}
where $\smash{E_m \equiv \sum_i \rho_i(a)/\rho_{\rm crit}^0}$ includes contributions of all matter and radiation fields, {\it i.e.} baryons, cold dark matter (CDM), photons and neutrinos, and $\smash{X(a)\equiv \rho_{\rm DE}^{\rm eff}(a)/\rho_{\rm DE}^{\rm eff}(a=1)}$ is due to any contribution to the standard Friedmann equation from terms other than the matter and radiation. Solving for the cosmological perturbations would require making additional assumptions regarding the underlying DE or modified gravity (MG) theory \cite{Silvestri:2009hh,Clifton:2011jh}, hence, only observables probing the background expansion were used in \citep{Wang:2018fng} to keep the reconstruction model-independent. The datasets included the CMB distance priors, the ``Joint Light-curve Analysis'' sample of supernovae type Ia (SNe Ia)~\citep{Betoule:2014frx}, the Hubble parameter $H_0$ from \citep{Riess:2016jrr}, the Observational Hubble parameters Data \citep{Moresco:2016mzx}, and the BAO distance measurements from i) the 6dF Galaxy Survey \citep{6df}, ii) SDSS DR7 Main Galaxy Sample \citep{MGS}, iii) the tomographic BOSS DR12 \citep{Zhao:2016das, Wang:2016wjr}, iv) eBOSS DR14 quasar sample \citep{Ata:2017dya} and the Lyman-$\alpha$ forest of BOSS DR11 quasars \citep{Font-Ribera:2013wce, Delubac:2014aqe}.

The correlated prior method assumes that, at each scale factor $a$, $X(a)$ is a Gaussian random variable with values at different $a$ correlated according to a specified correlation function, taken to be \citep{2012JCAP...02..048C, Crittenden:2005wj},
\begin{equation}
\xi(|\Delta a|) = \frac{\xi(0)}{1+(|\Delta a| / a_c)^2}.
\end{equation} 
As demonstrated in  \citep{2012JCAP...02..048C}, the details of the particular functional form are not essential as long as it has the rough shape that interpolates from 1 at $\Delta a=0$ to 0 at $\Delta a\gg a_c$. Here, $a_c$ determines the correlation length and $\xi(0)$ sets the strength of the prior and is related to the expected variance of the mean $\sigma_{\bar{X}}^2$ through $\sigma_{\bar{X}}^2 \simeq \pi \xi(0)a_c / (a_{\rm max} - a_{\rm min})$. The Gaussian prior effectively acts as an extra term in the total $\chi^2$, that is used to constrain the values of $X(a)$ in $40$ bins in the interval $a\in[0.001,1]$. The advantage of this approach is that it allows one to control the strength of the prior and find the Bayesian evidence for each choice of the prior parameters. If the evidence for DE dynamics is larger than that for $\Lambda$CDM for a broad range of values of $a_c$ and $\sigma_{\bar{X}}$, {\it i.e.} does not require one to optimize them to improve the evidence, then one could say that dynamical DE is favoured by observations. One can also define the evidence-weighted reconstruction, in which departures from $X(a)=1$ with low evidence get suppressed (see \citep{Wang:2018fng} for details).

Fig.~\ref{fig:xDE_wDE_rec} shows the DE density reconstruction performed with the ``standard'' choice of the prior, $\sigma_{\bar{X}}\equiv 0.04$, $a_c=0.06$ (in green) along with the evidence-weighted reconstruction (in blue). They show two apparent trends: an overall increase in the effective DE density and an oscillatory behaviour at $a\gtrsim 0.6$. The increase is driven by the local measurements of the Hubble constant $H_0$, whose larger value could be interpreted as an increase in DE density. The measurement of the BAO scale from the Lyman-$\alpha$ forest, which prefers a lower $H(z)$ at $z\sim 2.3$, further contributes to the same trend\footnote{The reconstruction in \citep{Wang:2018fng} was based on the Ly$\alpha$BAO analysis of BOSS DR11 quasars \citep{Delubac:2014aqe} that showed a $2.5\sigma$ deviation from the best fit $\Lambda$CDM. The tension has since been reduced to $1.7\sigma$ with the eBOSS DR14 Ly$\alpha$BAO analysis performed in \cite{Agathe:2019vsu,Blomqvist:2019rah}. Note that, although the tension is lower for the eBOSS sample, which has roughly 20\% more Lyman-$\alpha$ sources than BOSS, it does not necessarily mean that the tension will further decrease with a larger sample. The source of the tension, which could be new physics or a yet unknown systematic error, remains unknown, and forthcoming experiments, such as DESI, will help to clarify the issue.}. Oscillations, on the other hand, are caused by the combination of the tomographic BAO and the JLA SNe Ia data which happen to have matching oscillatory patterns.

One can see that the apparently large deviation from $X(a)=1$ at high redshifts, seen in the ``standard'' reconstruction in Fig.~\ref{fig:xDE_wDE_rec}, is not present in the evidence-weighted curve. The ability of data to constrain DE at $z>3$ is very weak and the reconstruction there is almost completely determined by the prior. This implies no Bayesian evidence for large deviations at high $z$, although the data still prefers a modest increase in DE density. 

The lower panels in Fig.~\ref{fig:xDE_wDE_rec} show the corresponding effective DE EOS $w_{\rm DE}^{\rm eff}(a)$. They are obtained by generating an ensemble of $X(a)$ from its mean and the covariance matrix and, for each realization, evaluate $w_{\rm DE}^{\rm eff}(a)$ from the conservation of the effective DE fluid. Averaging over the ensemble gives the mean and the uncertainty in $w_{\rm DE}^{\rm eff}(a)$ shown in the plots. If a sampled $X(a)$ happens to have a $|X(a)|<10^{-5}$, we replace it with $X(a)=10^{-5}$ to prevent a singularity in $w_{\rm DE}^{\rm eff}$. As expected, the uncertainty in $w_{\rm DE}^{\rm eff}$ is very large at high redshifts in the case of the standard prior (left panel). This is because $w_{\rm DE}^{\rm eff}(a)$ is determined by the derivative of $X(a)$ and each sampled $X(a)$ can fluctuate within the range allowed by the variance. In the case of the evidence-weighted reconstruction (right panel), $X(a)$ is a linear superposition of many reconstructions obtained with different priors. The different priors all prefer $X(a)$ to be constant, without enforcing any particular value of the constant. Thus, while there is $\sim$15\% uncertainty around the value of $X(a)$ at high $z$, its derivative is zero with a much higher certainty, which explains why the uncertainty in $w_{\rm DE}^{\rm eff}(a)$ is so small near $a=0$.

Interestingly, as shown in \citep{Wang:2018fng}, the Bayesian evidence ($\Delta \ln E$) for the oscillatory features is positive at $2.8\sigma$, and they appear equally prominently in both reconstructions in Fig.~\ref{fig:xDE_wDE_rec}. We also note that, although the Bayesian evidence for dynamical DE is weak, it has increased over the years, with the dynamical pattern being largely consistent with the reconstruction performed in 2012 \cite{Zhao:2012aw}. 

As we will see later, the oscillatory features in the reconstructions can, in certain circumstances, trigger fast-growing instabilities in cosmological perturbations. Also, the oscillatory pattern and the overall increase in DE density are driven by entirely different datasets.
For this reason, we have also considered $X(a)$ obtained by fitting a monotonic function to the reconstructions, which capture the overall increase but do not allow for oscillations. We take the form to be
\begin{equation}
\label{eqn:xDE_fitting_function}
X_{\rm fit}(a) = A \, \tanh [B(a-C)] + D,
\end{equation}
where the parameter $D$ is chosen such that $X_{\rm fit}(a=1) = 1$. The fitted functions and the corresponding DE EOS are shown with black solid and dashed lines in Fig.~\ref{fig:xDE_wDE_rec}. Thus, in what follows, we will consider four $X(a)$ histories: 
\begin{itemize}[label={}]
\item H1: using the standard prior (green line, Fig.~\ref{fig:xDE_wDE_rec});
\item H1F: the monotonic fit to H1 (black solid line);
\item H2: evidence-weighted reconstruction (blue line);
\item H2F: the monotonic fit to H2 (black dashed line).
\end{itemize}

As an increasing effective DE density cannot be realized in simple quintessence models, one is prompted to consider more complex gravity theories. In the next section we explore the GBD theories as a possible framework for explaining the observed DE dynamics. 

\section{Generalized Brans-Dicke theories and ways to reconstruct them}
\label{sect:GBDtheories}

The non-minimal coupling of the scalar field in the GBD theories could explain the observed ``ghostly'' behaviour of the effective DE density. We stress that, in this context, the phantom dynamics is only an apparent phenomenon perceived by a cosmologist fitting the conventional Friedmann equation to data while being unaware of the non-minimal coupling. 

The GBD action can be written as \citep{Bergmann1968, Nordtvedt1970, Wagoner1970}
\begin{equation}
\begin{split}
S & = \int d^4x \sqrt{-g} \left[ \frac{m_0^2}{2} F(\phi) R - \frac{1}{2} K(\phi) (\partial \phi)^2 - U(\phi) \right] \\
& + S_m[g_{\mu \nu}, \chi_i],
\end{split}
\label{eqn:GBDaction}
\end{equation}
where $m_0 \equiv (8\pi G)^{-1/2}$ is the Planck mass in terms of the Newton's constant $G$ measured on Earth, $\phi$ is the extra scalar degree of freedom, $(\partial \phi)^2 \equiv g^{\mu \nu} \partial_{\mu} \partial_{\nu} \phi$, $U(\phi)$ is the GBD potential and $S_m$ denotes the action for the matter fields $\chi_i$ minimally coupled to the (Jordan frame) metric $g_{\mu \nu}$. We set $K(\phi)=1$, as one can always do so by a redefinition of $\phi$. The modified Einstein equations are obtained by varying the action with respect to (w.r.t.) the metric $g_{\mu \nu}$,
\begin{equation}
\label{eqn:EinsteinEqGBD}
F G_{\mu \nu} = \frac{1}{m_0^2} \left( T_{\mu \nu}^{\mathrm{m}} +   T_{\mu \nu}^{\phi}\right) + \nabla_{\mu} \nabla_{\nu} F - g_{\mu \nu} \square F,
\end{equation}
where $\nabla_{\mu}$ denotes the covariant derivative w.r.t. the coordinate $x^{\mu}$, $\square \equiv g^{\mu \nu} \nabla_{\mu} \nabla_{\nu}$, $T_{\mu \nu}^{\rm m}$ is the matter energy-momentum tensor and
\begin{equation}
T_{\mu \nu}^{\phi} \equiv  \partial_{\mu} \phi \partial_{\nu} \phi - g_{\mu \nu} \left[ \frac{1}{2} \partial_{\alpha} \phi \partial^{\alpha} \phi + U(\phi)\right].
\end{equation}
The equation of motion for the scalar field $\phi$ is then obtained by extremizing the action  \eqref{eqn:GBDaction} w.r.t. variations of the field $\phi$,
\begin{equation}
\square \phi = U_{\phi} - \frac{m_0^2}{2} F_{\phi} R,
\label{eq:boxphi}
\end{equation}
where the subscript $\phi$ denotes a derivative w.r.t. $\phi$.
For convenience we redefine the field, $\smash{ \phi \to \phi / m_0}$, to make it dimensionless, and the potential, $\smash{U \to U / m_0^2}$, with the latter measured in ${\rm Mpc}^{-2}$ in agreement with the units convention in {\tt CAMB} \citep{Lewis:1999bs}.

The freedom in choosing the two functions $F(\phi)$ and $U(\phi)$ translates into the ability of GBD theories to reproduce any expansion history. In particular, the effective DE density $X(a)$ defined in Eq.~(\ref{eqn:FriedmannPhenomen}) can increase, and the effective DE EOS can cross $-1$. To see this, we re-write the modified Einstein equation (\ref{eqn:EinsteinEqGBD}) as
\ba
\nonumber
G_{\mu \nu} &=&  \frac{1}{m_0^2 F} \left\{ T^m_{\mu \nu} + T^\phi_{\mu \nu} + \nabla_\mu \nabla_\nu F- g_{\mu \nu} \Box F \right\} \\
&=& \frac{1}{m_0^2}  \left\{ T^M_{\mu \nu} + (T^{\rm eff}_{\rm DE})_{\mu \nu} \right\}\,,
\label{eq:modeinstein}
\ea
where, in the second line, we have defined the effective DE stress-energy by absorbing into it all the terms on the right hand side other than the usual matter term, {\it i.e.},
\be
\nonumber
(T^{\rm eff}_{\rm DE})_{\mu \nu} \equiv F^{-1} \left\{ T^\phi_{\mu \nu} + \nabla_\mu \nabla_\nu F- g_{\mu \nu} \Box F + (1-F) T^m_{\mu \nu} \right\}\,.
\ee
In a flat FRW Universe, the effective DE density is 
\be
\rho^{\rm eff}_{\rm DE} = F^{-1} \left\{\dot{\phi}^2/(2a^2) + U(\phi) - 3\mathcal{H} \dot{F}/a^2 +(1-F)\rho_m \right\}\,,
\label{eq:rhoeff}
\ee
with the dot standing for a derivative w.r.t the conformal time, while the effective DE pressure is
\be
p^{\rm eff}_{\rm DE} = F^{-1} \left\{\dot{\phi}^2/(2a^2)  - U(\phi) +\mathcal{H}\dot{F}/a^2 +\ddot{F}/a^2 \right\}\,.
\ee
The $\mu=\nu=0$ component of Eq.~(\ref{eq:modeinstein}) gives the Friedmann equation,
\be
\mathcal{H}^2 = \left( \dot{a} \over a \right)^2 = {a^2 \over 3m_0^2} [ \rho_m(a) + \rho^{\rm eff}_{\rm DE}(a)],
\ee
which can be recast in the form of Eq.~(\ref{eqn:FriedmannPhenomen}). 

Note that, by construction, the effective DE ``fluid'' is conserved, but its EOS,
\be
w^{\rm eff}_{\rm DE}= {\dot{\phi}^2/(2a^2) - U(\phi) +\mathcal{H}\dot{F}/a^2+\ddot{F}/a^2 \over \dot{\phi}^2/(2a^2) + U(\phi)- 3\mathcal{H} \dot{F}/a^2 +(1-F)\rho_M}\,,
\ee
is not always well-defined because $\rho^{\rm eff}_{\rm DE}$ in the denominator is allowed to change sign due to the new terms generated by the non-minimal coupling $F(\phi)$. Thus, as previously noted in \citep{Carroll:2004hc,Das:2005yj}, observing $\smash{w^{\rm eff}_{\rm DE}<-1}$, or finding that $\rho^{\rm eff}_{\rm DE}$ changes its sign, could be a smoking gun of interactions in the dark sector. 

The idea of reconstructing the GBD Lagrangian from a given expansion history was previously explored in \citep{Boisseau:2000pr, EspositoFarese:2000ij, Perivolaropoulos:2005yv}, motivated by the fact that the Hubble function $H(a)$ inferred from the supernovae data available at that time showed \gb{a} preference for an effective phantom DE equation of state, $\smash{w^{\rm eff}_{\rm DE} < -1}$. As they have shown, one can, in principle, reconstruct both functions $F(\phi)$ and $U(\phi)$ if, in addition to $H(a)$, one knows the evolution of the growth of the matter density contrast $\delta(a)$. Another interesting example is the $f(R)$ gravity where the only unknown function is the function $f$ itself and the full reconstruction can be done with the sole knowledge of the expansion history $H(a)$~\citep{Song:2006ej,Pogosian:2007sw}.

In the present work we adopt a slightly different approach. Since the growth of perturbations is rather complicated to extract in a model-independent way because of the redshift-space distortions, non-linearities, bias, \gb{\etc}, we attempt to reconstruct only one of the functions, namely $U(\phi)$, while the other is chosen to either have a given functional form $F(\phi)$ (Model 1), or a given parameterized time-dependence $F(a)$ (Model 2). We will analyze these two cases separately.

While exploring the parameter space, which includes the initial conditions for the scalar field, we restrict to solutions in which the net change in $F(\phi)$ is under 10\%, to satisfy the BBN constraints on the variation of the Newton's constant. We also check for various types of instabilities using the procedure implemented in {\tt EFTCAMB} \citep{Hu:2013twa,Raveri:2014cka}. Specifically, we check for ghost, gradient and mass instabilities discussed in detail in \cite{Frusciante:2018vht} and briefly reviewed below.

After expanding the action up to the second order in perturbations of the metric and matter fields, and removing spurious degrees of freedom, one can isolate the action for the propagating scalar and tensor degrees of freedom~\cite{DeFelice:2016ucp}. The conditions for avoiding instabilities can then be formulated in Fourier space in terms of the corresponding kinetic, gradient and mass matrices as follows:\begin{enumerate}
\item \textit{no-ghost:} a ghost instability develops when the kinetic term of a field is negative. In the presence of multiple propagating degrees of freedom, a positive definite kinetic matrix guarantees that no ghosts will develop. In practice, this requirement needs to be imposed only at high energies, \gb{\ie}~ in the high-$k$ limit, since an infrared ghost does not lead to catastrophic instabilities~\cite{Gumrukcuoglu:2016jbh}.
\item \textit{no-gradient:} gradient instabilities arise in the high-$k$ regime when the speed of propagation is imaginary. The sound speeds of the propagating degrees of freedom can be identified from the dispersion relations that result from the quadratic action after diagonalizing the kinetic matrix. In order to avoid gradient instabilities we impose $c_s^2>0$ for all the degrees of freedom. 
\item \textit{no-tachyon}: whenever the mass matrix of the Hamiltonian contains a negative eigenvalue, the mass instability plagues the low-$k$ regime with the development of a tachyon~\cite{DeFelice:2016ucp}. The rate of the instability needs to be taken into account. We will assume that the GBD Hamiltonian exhibits a tachyonic instability when at low momenta a mass eigenvalue $\mu_i$ becomes negative and evolves rapidly, \gb{\ie}~$|\mu_i| \gg H^2$. Thus, for a theory to be viable, we require $\mu_i>0$ or, alternatively, $|\mu_i|\lesssim H^2$. 
\end{enumerate}
This set of conditions was shown to guarantee stability over the full range of linear scales \cite{Frusciante:2018vht} and was implemented in a private version of {\tt EFTCAMB}. 

Public versions of {\tt EFTCAMB}, as well as other Einstein-Boltzmann solvers like {\tt HiClass} \cite{Zumalacarregui:2016pph}, do not contain the mass condition. Instead, in addition to checking for the no-ghost and no-gradient instabilities, they impose a set of mathematical conditions that prevent the development of exponentially diverging solutions. 
The latter are worked out from the linear order equation for the scalar field perturbation and are meant to protect against the mass instabilities as well as the ghost and gradient instabilities that could have possibly evaded the checks based on some approximations necessary in setting the conditions. When a mathematical condition is violated, one cannot easily tell which of the three types of instabilities was responsible.
 In our analysis we used both methods. Namely, we checked for the ghost, gradient and mass stability conditions, as well as using the publicly available stability check that combines the ghost, the gradient and the mathematical conditions.

\subsection{Model 1: Reconstructing GBD for a given $F(\phi)$}
\label{sec:designer_gbd_fphi}

Given the functional form of $F(\phi)$, we can reconstruct $U(\phi)$ from a given expansion history. We take 
\begin{equation}
\label{eqn:GBDcoupling}
F(\phi) = \exp(\xi \phi),
\end{equation}
which is a form motivated by high energy theories, \eg~a non-minimally coupled dilaton field representing compactified extra dimensions with the dimensionless parameter $\xi$ controlling the coupling strength. 

We begin by writing the two Friedmann equations as
\begin{align}
\label{eqn:1stFriedmannGBD}
\mathcal{H}^2 & = \frac{1}{\cal D}\frac{ \rho \, a^2}{3 m_0^2} +   \frac{1}{\cal D} \frac{ Ua^2}{3}, \\
\label{eqn:2ndFriedmannGBD}
\begin{split}
\mathcal{G} \frac{\ddot{a}}{a} & =  2 \mathcal{D} \mathcal{H}^2 - \frac{a^2 \mathcal{H}^2}{2} \left( \frac{1}{3} + F_{\phi \phi} \right) (\phi^{\prime})^2  \\
&   - \frac{1}{2} \frac{(\rho + P)a^2}{m_0^2} - \frac{1}{2} F_{\phi}  \mathcal{H}^2 \left(\phi^{\prime \prime} - \phi^{\prime}\right),
\end{split}
\end{align}
where
\begin{align}
\label{eqn:functionF}
{\cal D} & = F - \frac{1}{6} (\phi^{\prime})^2 +  F_{\phi} \phi^{\prime}, \\
{\cal G} & = F +\frac{1}{2}  F_{\phi} \phi^{\prime},
\end{align}
and ${}^{\prime}$ denotes derivatives w.r.t. $N\equiv\ln a$. Eq.~\eqref{eqn:2ndFriedmannGBD} can be rewritten as an equation for the background evolution of $\phi$:
\begin{equation}
\label{eqn:EqnGBDPhi}
\begin{split}
\phi^{\prime \prime} & = -\frac{1+ F_{\phi \phi}}{F_{\phi}} (\phi^{\prime})^2  \\
& + \left(1 + \frac{1}{2} \frac{3 E_m + 4 E_r - E_{\nu}^{\prime} - \Omega_{\Lambda} X^{\prime}}{E_m + E_r + E_{\nu} + \Omega_{\Lambda} X} \right) \phi^{\prime} \\
& + \frac{1}{F_{\phi}}\frac{(F-1) (3 E_m + 4 E_r - E_{\nu}^{\prime})- F \Omega_{\Lambda} X^{\prime}}{E_m + E_r + E_{\nu} + \Omega_{\Lambda} X},
\end{split}
\end{equation}
where $\smash{E_m \equiv \rho_m / \rho_{\rm crit}^0}$ includes CDM and baryons, $\smash{E_r \equiv \rho_r / \rho_{\rm crit}^0}$ includes  photons and massless neutrinos and $\smash{E_{\nu} \equiv \rho_{\nu} / \rho_{\rm crit}^0}$ includes massive neutrinos species only. Eq~\eqref{eqn:EqnGBDPhi} can be solved given the functional form \eqref{eqn:GBDcoupling} of $F(\phi)$ and the DE density evolution $X(a)$. Given the solution, $\phi(a)$, one can find the potential $U(a)$ from Eq.~\eqref{eqn:1stFriedmannGBD}, namely:
\begin{equation}
\label{eqn:GBD_reconstruct_potential}
\begin{split}
U a^2 & = 3 \mathcal{D} H_0^2 a^2 (E_m+E_r+E_{\nu}+\Omega_{\Lambda}X) \\
& -  3 H_0^2 a^2 (E_m+E_r+E_{\nu}).
\end{split}
\end{equation}
If $\phi(a)$ is monotonic, it can be inverted to obtain $a(\phi)$ and, thus, $U(\phi)$ for the range of $\phi$ covered by the evolution.

\begin{figure}[tbph]
\centering
\includegraphics[width=.45\textwidth]{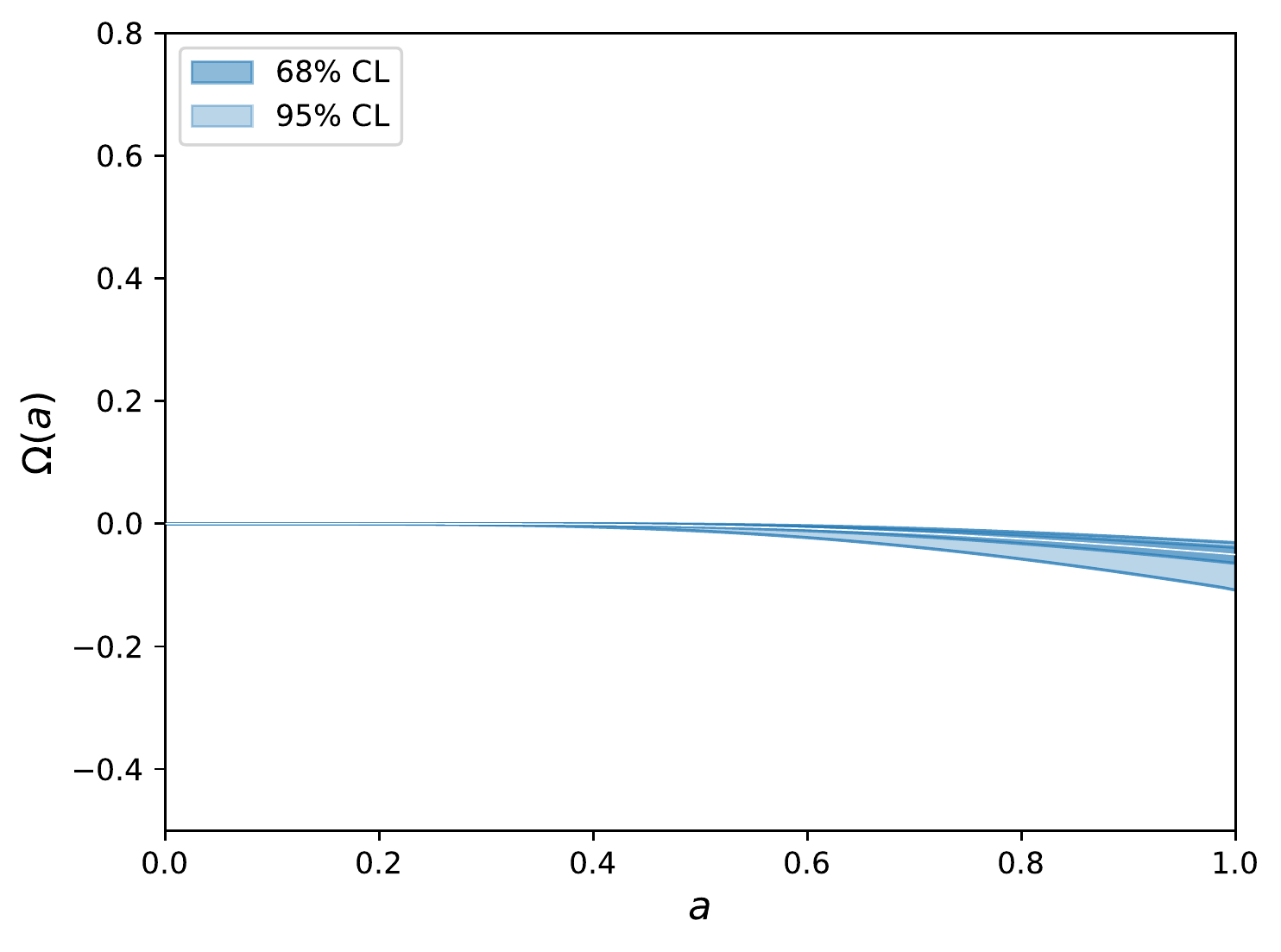}
\caption{\label{fig:GBD_wei_tanh_omega} The viable range of values of the non-minimal coupling $\Omega(a) \equiv F(a) - 1$ for the Model 1 GBD theories reconstructed from the H2F expansion history. The confidence level (CL) regions are obtained by sampling parameters $\phi^{\prime}_{\rm ini}, \xi$ and $\log_{10}a_{\rm ini}$ as described in the text.}
\end{figure}

\begin{figure}[tbph]
\centering
\includegraphics[width = .45\textwidth]{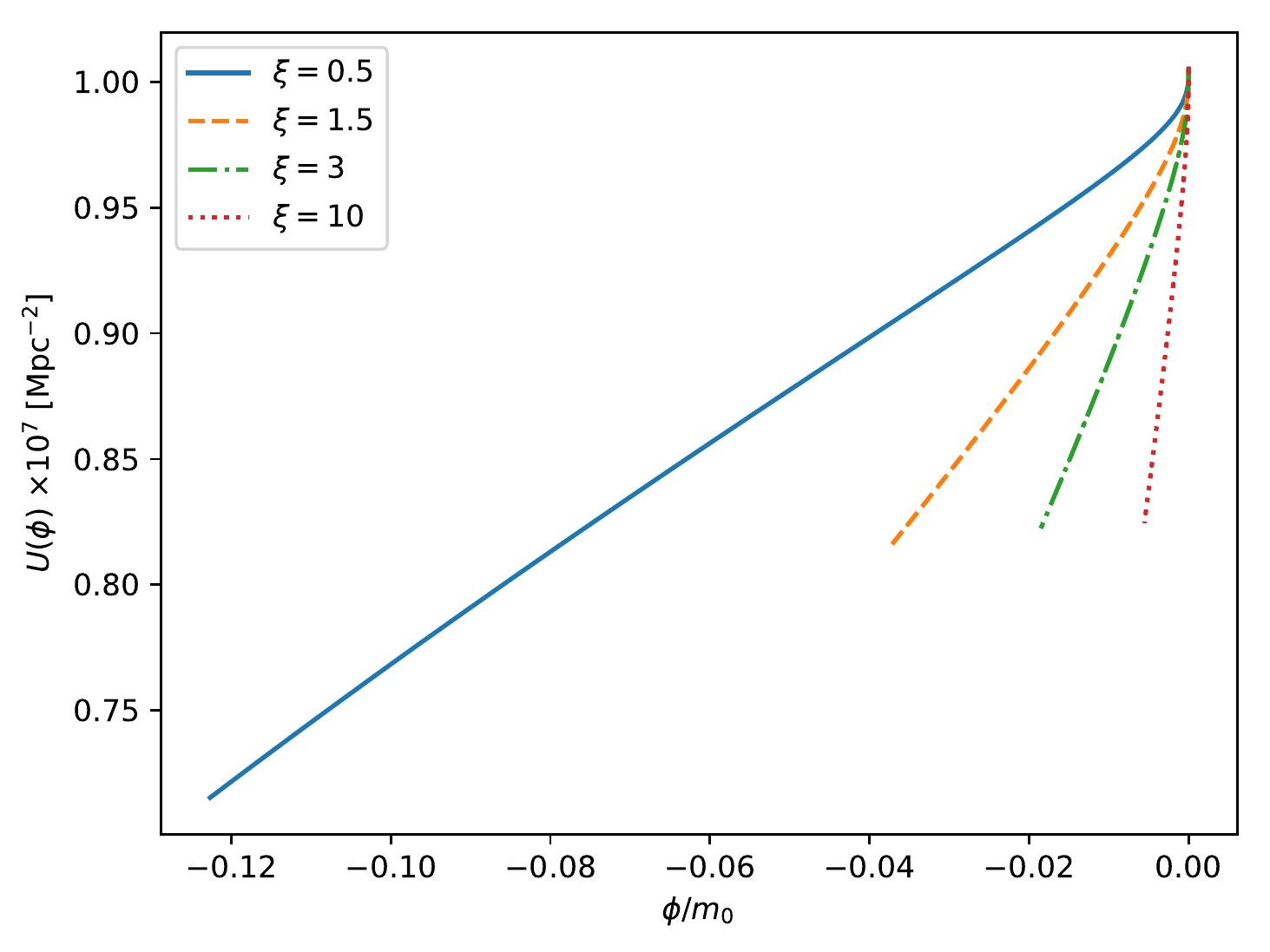}
\caption{\label{fig:GBD_model_potential} The potential $U(\phi)$ in four representative Model 1 GBD theories reconstructed from the H2F expansion history with the same $\phi^{\prime}_{\rm ini}$ and $\log_{10}a_{\rm ini}$, and our different values of the coupling parameter $\xi$.}
\end{figure}

Solving Eq.~\eqref{eqn:EqnGBDPhi} requires setting the value of the field $\phi_{\rm ini}$ and its derivative $\phi_{\rm ini}^{\prime}$ at some initial time $a_{\rm ini}$. To preserve the success of $\Lambda$CDM in explaining the Big Bang Nucleo-synthesis (BBN) and the peak structure of the CMB spectrum, we assume that gravity was close to GR at early times, so that $\smash{F(\phi) = 1}$ for $\smash{a \le a_{\rm ini}}$, but could start deviating from unity at later times. For $\smash{F(\phi) = \exp(\xi \phi)}$ this means $\smash{\phi_{\rm ini} = 0}$ and, to explain the features in the reconstructed DE density discussed in the previous section, we will need $\smash{a_{\rm ini} \lesssim 0.1}$. Thus, in addition to providing $X(a)$, we have to specify three parameters: $\xi$, $a_{\rm ini}$ and $\phi_{\rm ini}^{\prime}$.

For the H1, H1F and H2 background histories, reconstructed Model 1 theories contain fast-growing mode instabilities for all choices of initial conditions. For the perturbations around these backgrounds we find that both the mass and the mathematical conditions are not satisfied. It appears that the large rapid increases in $X(a)$ present in H1, H1F and H2 drive the solution towards instability, which could, in principle, be prevented by an appropriate choice of $F(\phi)$. However, the Model 1 coupling function $F(\phi) = \exp(\xi \phi)$ is monotonic and is unable to prevent the onset of instability.


In the case of H2F, which is monotonic and with a relatively small change in $X(a)$, we are able to find viable solutions despite finding negative mass eigenvalues. There, the tachyonic instability corresponding to the negative mass eigenvalues develops on time scales comparable to the Hubble rate allowing for growth of cosmic structure that is in reasonable agreement with observations.

To summarize, we find that Model 1 reconstructions from all four expansion histories are plagued by a mass instabilities. For H1, H1F and H2, this instability develops on time scales small enough for the mathematical condition to detect diverging solutions. In the case of H2F the characteristic time scale is longer and, while our approximate bound of $|\mu_i|<H^2$ is not satisfied, the instability does not develop to the point of giving diverging solutions. 

Fig.~\ref{fig:GBD_wei_tanh_omega} shows the allowed range of the non-minimal coupling function $\Omega(a) \equiv F(\phi(a))-1$ for GBD theories reconstructed from H2F. It is obtained by uniformly sampling parameters $(\phi^{\prime}_{\rm ini}, \log_{10}a_{\rm ini}, \xi )$ from the intervals
\begin{equation}
\label{eqn:GBD_sampling_prior}
\frac{\phi^{\prime}_{\rm ini}}{m_0} \in [-10^{-6}, 10^{-6}], \,\, \log_{10} a_{\rm ini} \in [-3,-1],\,\, \xi \in [0.1, 10],
\end{equation}
solving for the evolution of $\phi$ and selecting solutions that have $\Omega$ within the allowed range and satisfy the stability condition.
The shaded regions in Fig.~\ref{fig:GBD_wei_tanh_omega} indicate the confidence level (CL) for having a particular value of $\Omega$ at a given $a$, while the dark line in the middle shows the mean. Examining the numerical solutions, we find that, as expected, the increasing effective DE density drives the field to negative values, resulting in $F(\phi)<1$ and a larger $G_{\rm eff} \propto G/F(\phi)$.

For illustration, in Fig.~\ref{fig:GBD_model_potential} we show the potential $U(\phi)$ for four GBD theories reconstructed from the H2F DE density with $\phi^{\prime}_{\rm ini}/m_0 = 0$ and $a_{\rm ini} = 10^{-2}$ and $\xi=0.5$, $1.5$, $3$ and $10$ respectively. We can see how in all four cases the potential has a cusp at the origin. Stronger couplings leads to steeper potentials. Their shapes resemble the potential in chameleon-like models  \cite{Khoury:2003rn,Brax:2004qh}, $V(\phi) \propto |\phi|^{-n}$, although the dynamics here is completely different. In the chameleon model, the field tracks the minimum of the effective potential, with the coupling function $F(\phi)$ slowly increasing with the evolution. In our reconstructed theories, the field $\phi$ starts at the top of the cusp and rolls down the potential, with $F(\phi)$ decreasing as it rolls.

\subsection{Model 2: Reconstructing GBD with a parameterized $F(a)$}
\label{sec:designer_gbd_fa}

\begin{figure*}[tbph]
\centering
\subfigure{
\includegraphics[width = .45\textwidth]{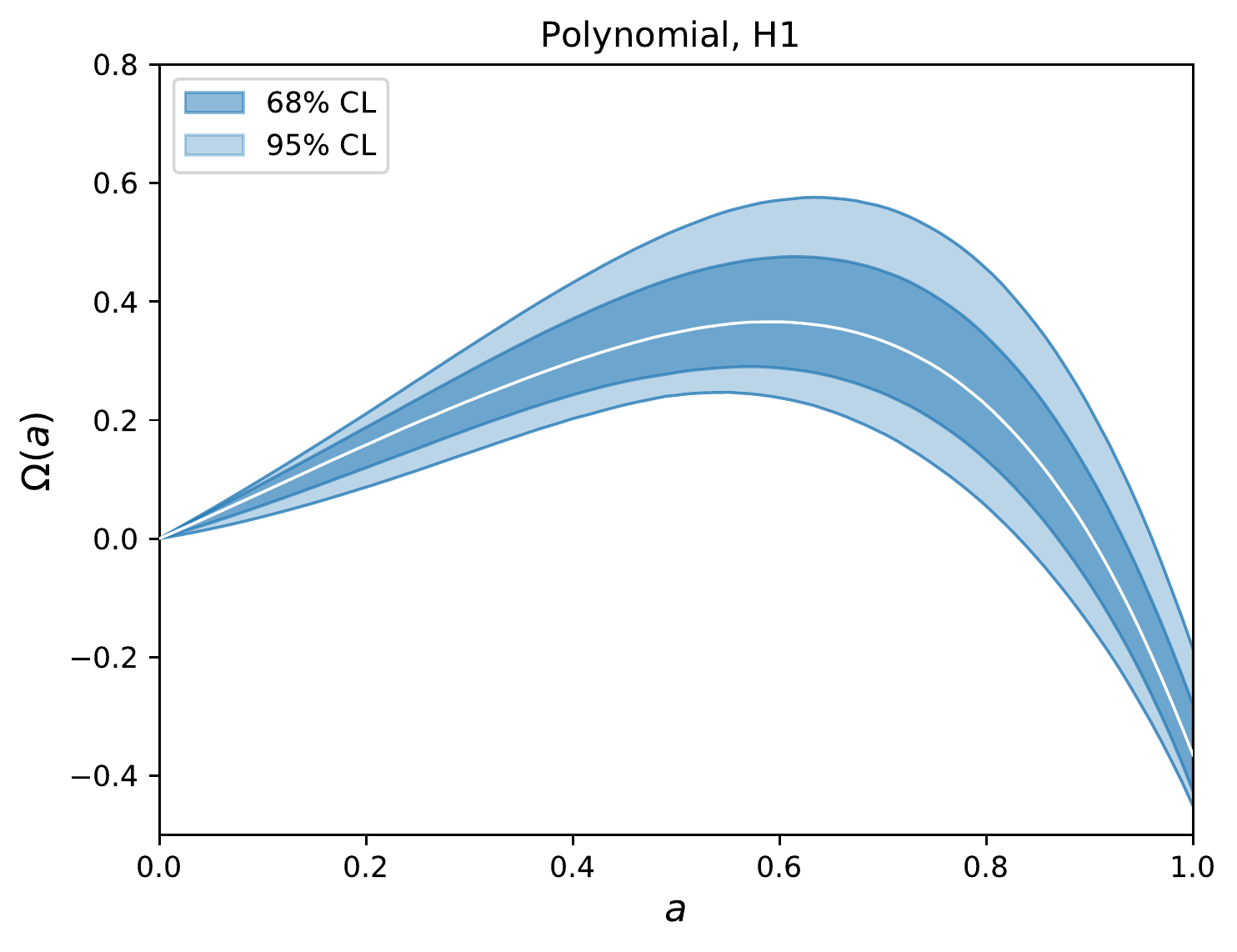}} \,
\subfigure{
\includegraphics[width = .45\textwidth]{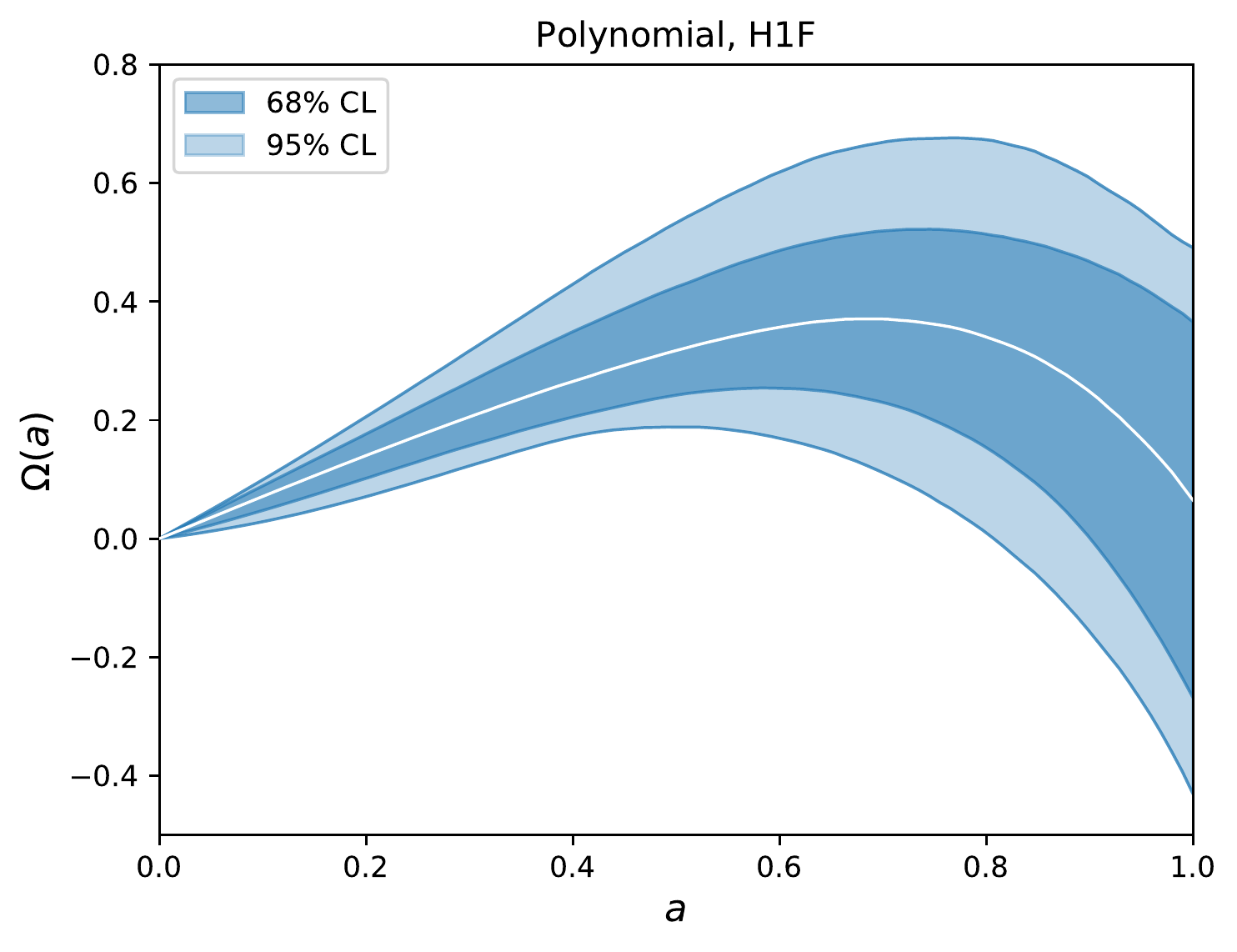}} \\
\subfigure{
\includegraphics[width = .45\textwidth]{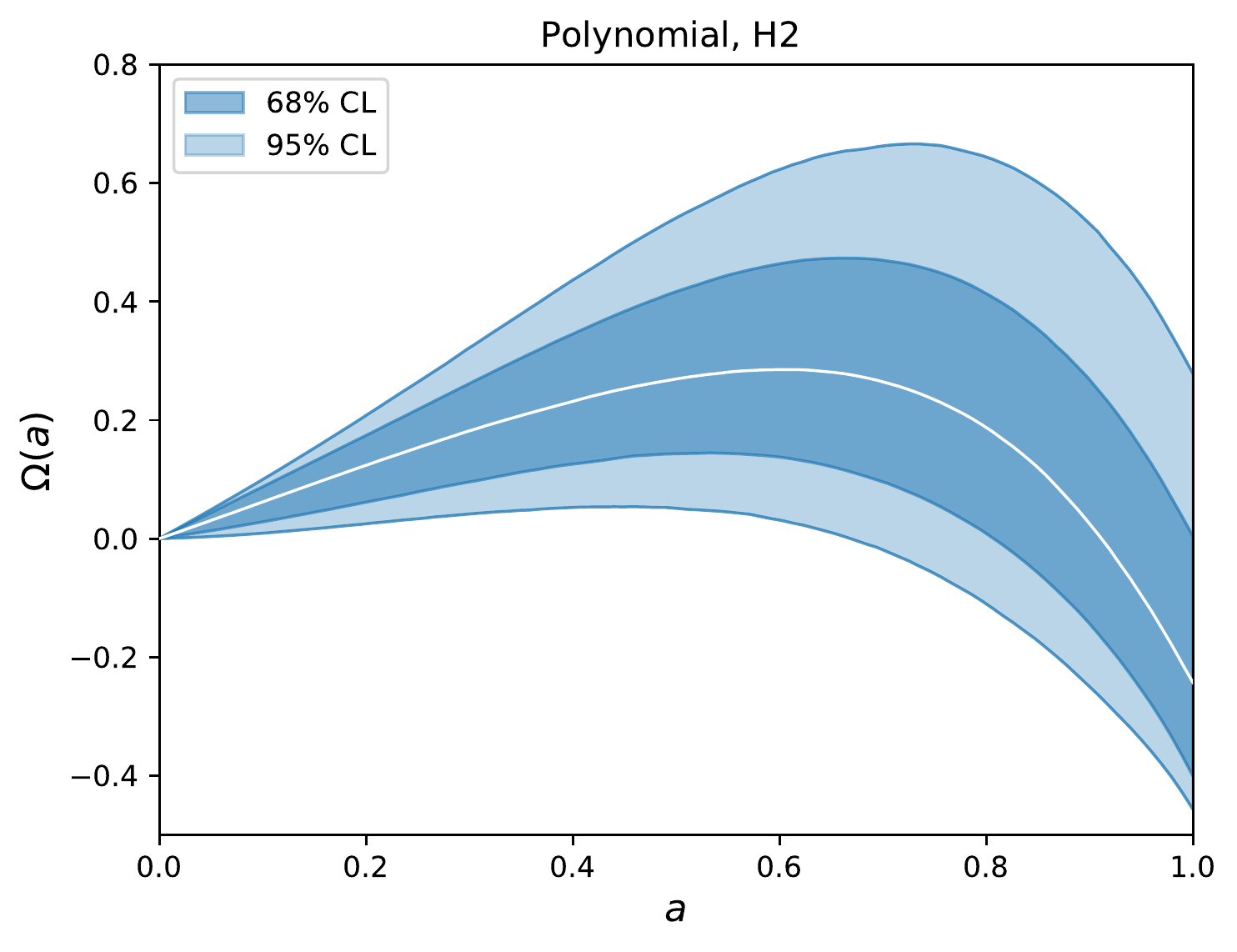}} \,
\subfigure{
\includegraphics[width = .45\textwidth]{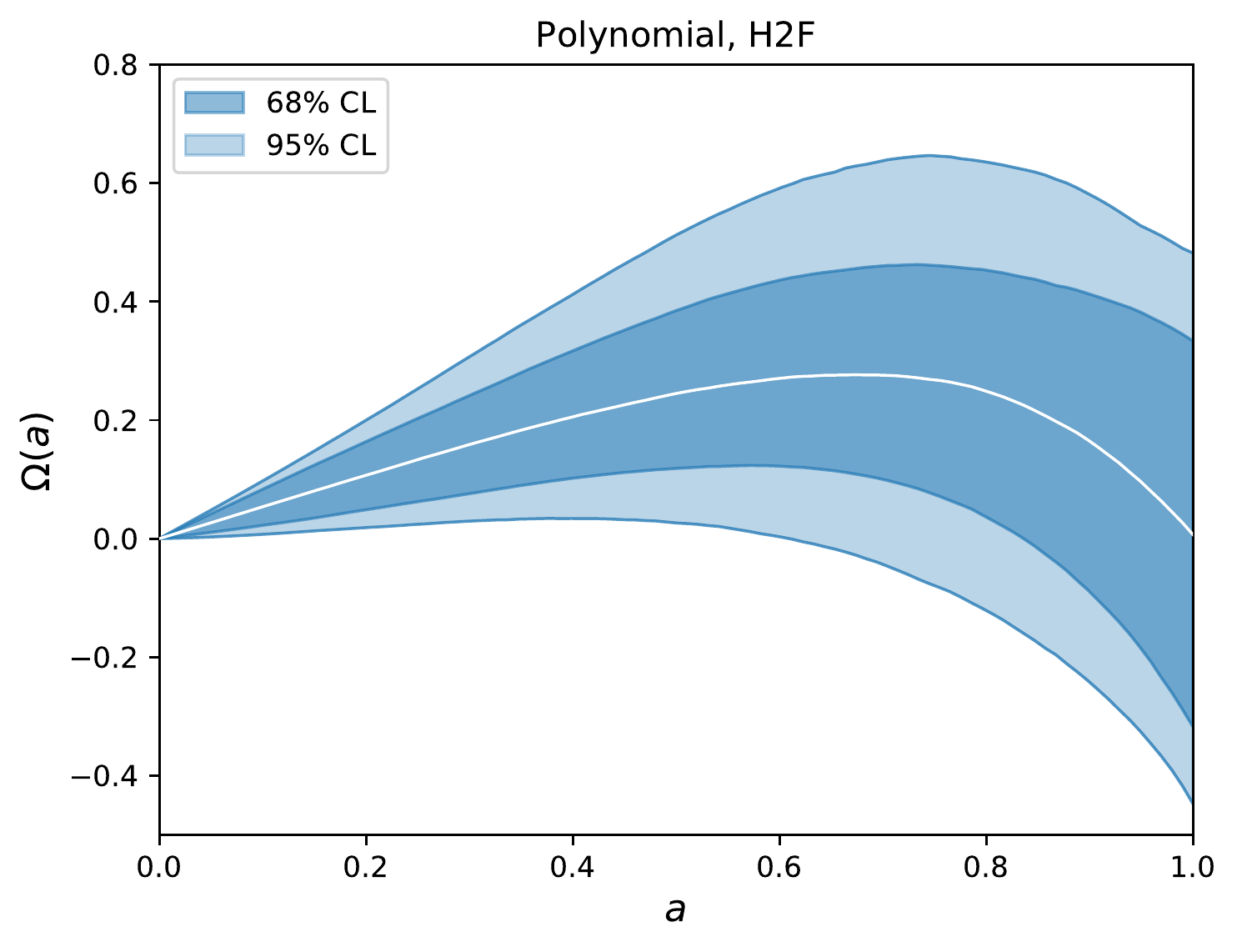}}
\caption{\label{fig:poly_omega} The distribution of non-minimal coupling functions $\Omega(a)=F(a)-1$ obtained using the Model 2 polynomial parametric form for stable GBD theories with the four effective DE histories H1, H1F, H2 and H2F. }
\end{figure*}

\begin{figure*}[tbph]
\centering
\includegraphics[width = .95\textwidth]{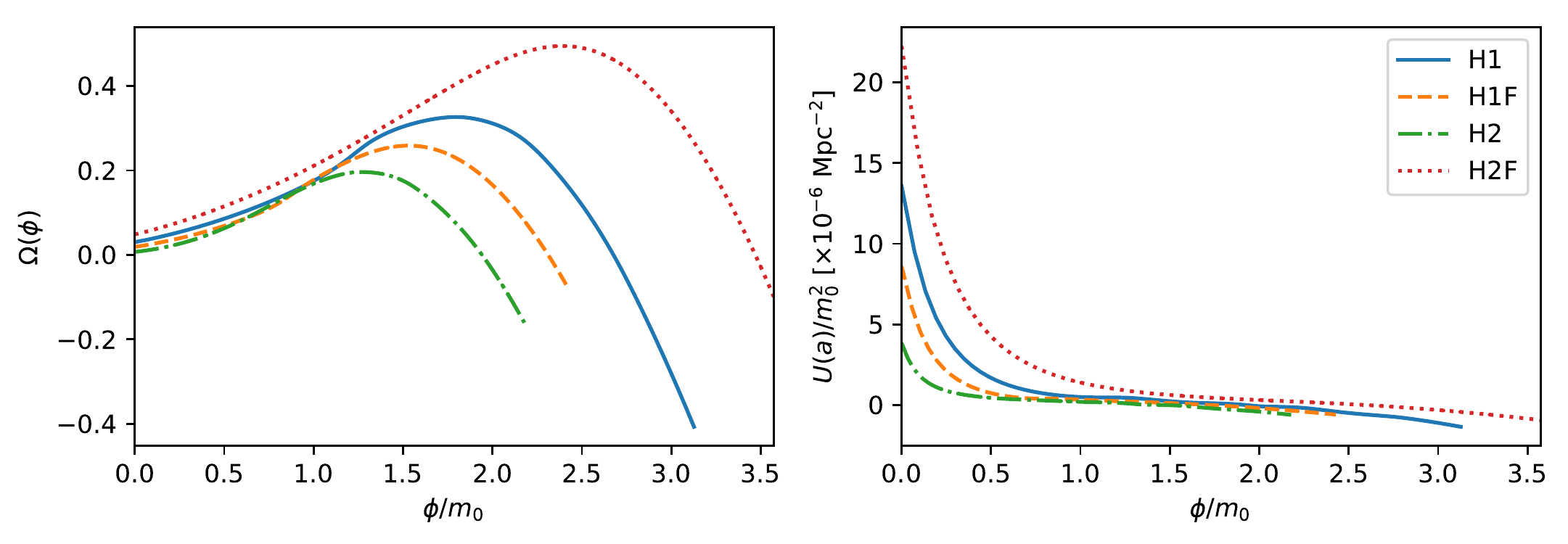}
\caption{\label{fig:reconstructed_gbd_poly} The coupling function $\Omega(\phi) = F(\phi)-1$ and the potential $U(\phi)$ for four representative Model 2 theories reconstructed from the H1, H1F, H2 and H2F expansion histories. }
\end{figure*}

We now change the approach and, instead of working with a given $F(\phi)$, we directly specify the time-dependence of $F$, {\it i.e.} $F(a)$. A similar approach was used in \citep{EspositoFarese:2000ij, Perivolaropoulos:2005yv} to reconstruct the GBD Lagrangian from the expansion history inferred from an early SNe Ia dataset. In this case, we start by writing the modified Friedmann equations as
\begin{align}
\label{eqn:GBDFirstFriedmann}
\mathcal{H}^2 & = \frac{1}{3 m_0^2} \frac{1}{F+aF^{\prime}} \left[ \rho a^2 + \frac{1}{2} \dot{\phi}^2 + Ua^2\right], \\
\label{eqn:GBDSecondFriedmann}
\begin{split}
\dot{\mathcal{H}} & = \frac{1}{F + \frac{1}{2}aF^{\prime}} \biggl\{ \left[F + 2aF^{\prime} + a^2 F^{\prime \prime} \right]  \mathcal{H}^2 \\
& - \frac{1}{2m_0^2} \left[ Pa^2 + \frac{1}{2} \dot{\phi}^2 - Ua^2 \right] \biggr\},
\end{split}
\end{align}
where primes denote derivatives w.r.t. the scale factor and overdots denote derivatives w.r.t. the conformal time $\tau$. We can then use \eqref{eqn:GBDFirstFriedmann} to eliminate the potential $U$ in \eqref{eqn:GBDSecondFriedmann} to write
\begin{equation}
\label{eqn:GBDSecondFriedmann_2}
\begin{split}
\dot{\mathcal{H}} & = \left\{ \left[ \frac{5}{2} F + \frac{7}{2}aF^{\prime} + a^2 F^{\prime \prime} \right]  \mathcal{H}^2 - \frac{(\rho+P)a^2}{2m_0^2} - \frac{1}{2m_0^2} \dot{\phi}^2 \right\} \\
& \times \left( F + \frac{1}{2}aF^{\prime} \right)^{-1}.
\end{split}
\end{equation}
One can then solve the above equation for $\dot{\phi}$ and use it in \eqref{eqn:GBDFirstFriedmann} to obtain a solution for $U(a)$:
\begin{equation}
\begin{split}
\frac{U a^2}{m_0^2} & = \mathcal{H}^2 \left[ \frac{1}{2} F - \frac{1}{2} a F^{\prime} - a^2 F^{\prime \prime} \right] \\
& + \frac{(P - \rho)a^2}{2 m_0^2} + \dot{\mathcal{H}} \left[ F + \frac{1}{2} a F^{\prime}\right] .
\end{split}
\label{eq:Ua}
\end{equation}
With the known $U(a)$, one can solve for the kinetic energy $\dot{\phi}^2$ from \eqref{eqn:GBDFirstFriedmann} and complete the solution by solving the differential equation to find $\phi(a)$.
With the field $\phi(a)$ known, one can convert $U(a)$ and $F(a)$ into $U(\phi)$ and $F(\phi)$, thus reconstructing the functional form of the theory for the range of $\phi(a)$ covered by the solution.

To explore a broad range of possible $F(a)$ histories we adopt a polynomial parametric form
\begin{align}
F(a) & = 1 + \sum_{i=1}^5 \alpha_i a^i,
\label{eq:Fa}
\end{align}
with coefficients $\alpha_i$ sampled uniformly from 
\begin{gather}
\label{eqn:parameter_space_poly}
\alpha_i \in [-1,1].
\end{gather}
This range is chosen to favour positive values of $F(a)$ close to unity as required by existing bounds.

With $H(a)$ and $F(a)$ specified, one can use {\tt EFTCAMB}  \citep{Hu:2014oga}, as described in the next Section, to compute the cosmological observables. 

We have generated samples of $F(a)$ using the parameterized form \eqref{eq:Fa} and performed reconstructions of the GBD theories for each of the four $X(a)$ histories shown in Fig.~\ref{fig:xDE_wDE_rec}. The viable ranges of $F(a)$ functions in each case are shown in Fig.~\ref{fig:poly_omega}. One can see that for H1, H1F and H2F, in which $X(a)$ has a large increase or is non-monotonic, $F(a)$ must be non-monotonic to avoid instabilities. In the case of H2F, which has a gently increasing monotonic $X(a)$, a monotonic $F(a)$ has a small probability, but is not excluded.

To reconstruct $U(\phi)$ and $F(\phi)$ one needs to solve a first order ODE for $\phi$, which requires specifying a boundary condition, such as the value of the field at an initial time $a_{\rm ini}$. This means we can only reconstruct $U(\phi)$ and $F(\phi)$ up to an arbitrary shift in the value of $\phi$. The shift has no physical significance, as all the observables are already fully determined. Hence, without loss of generality, we take $a_{\rm ini}=0.001$ and $\phi(a_{\rm ini}) = 0$. 

Fig.~\ref{fig:reconstructed_gbd_poly} shows the non-minimal coupling function $\Omega(\phi)=F(\phi)-1$ and the potential $U(\phi)$ for four representative theories reconstructed from H1, H1F, H2 and HF2.  We see that $\Omega(\phi)$ is non-monotonic in these representative cases. The potentials have a runaway shape, being seemingly unbounded from below for large values of the field, although one should keep in mind that the shape is only known over the range covered by the evolution of the field. One can also see small bumps in the potentials derived from H1 and H2, needed to accommodate oscillations in $X(a)$.

\section{Cosmological Observables in reconstructed GBD theories}
\label{sec:observables}

\begin{figure*}[tbph]
\centering
\subfigure{
\includegraphics[width=.45\textwidth]{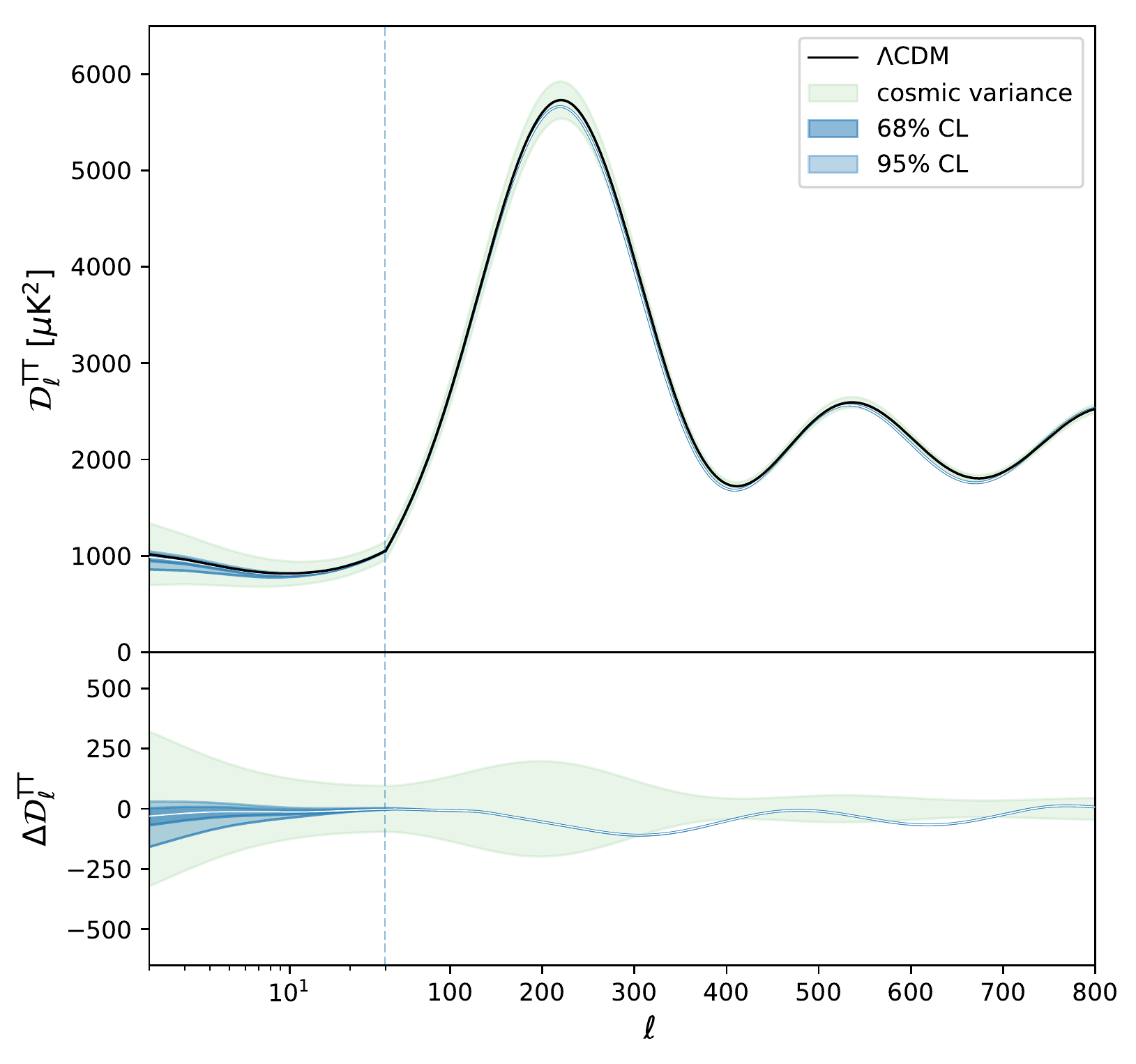}} \,
\subfigure{
\includegraphics[width=.45\textwidth]{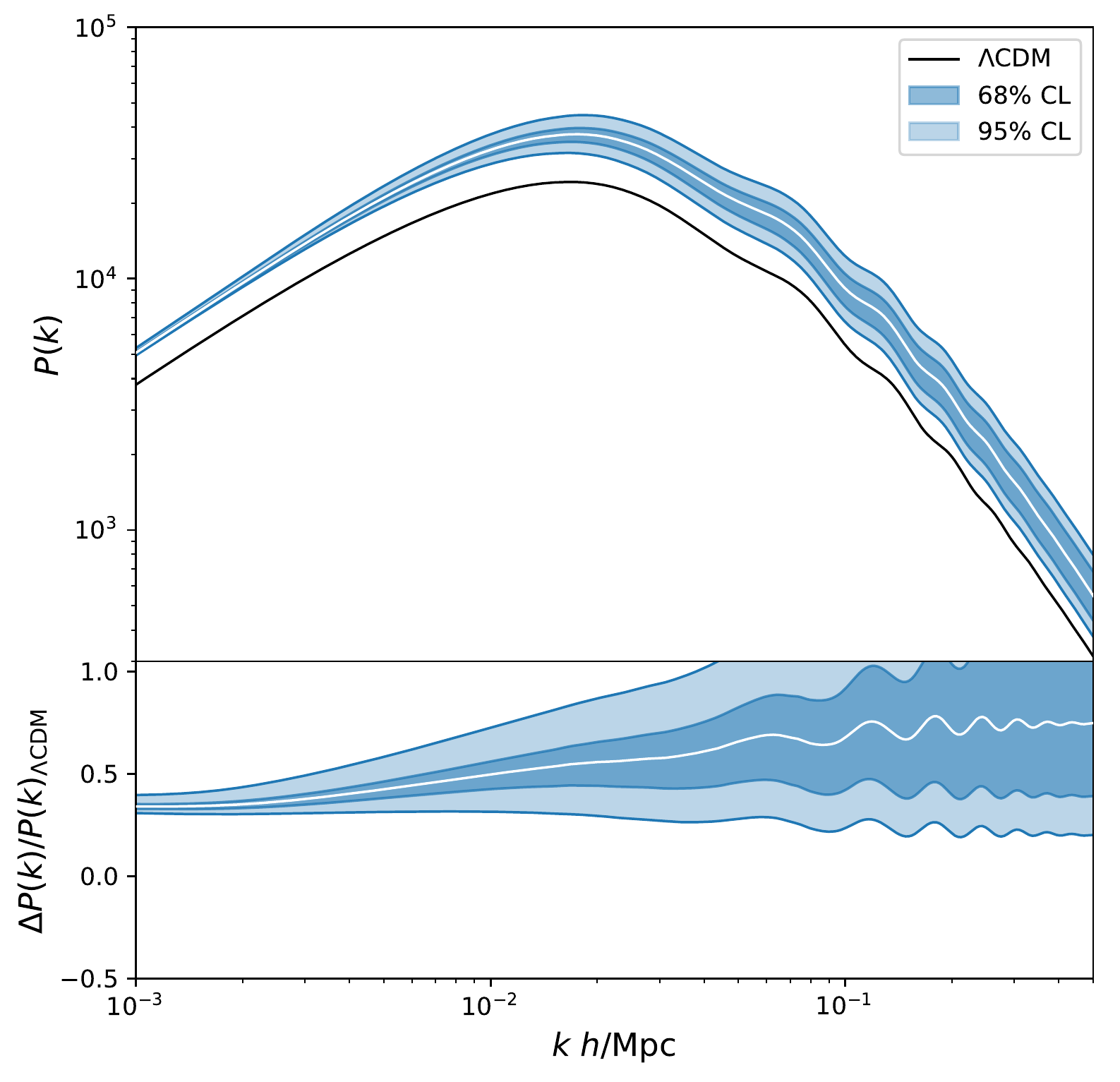}}
\caption{\label{fig:GBDimpact_CMB_MPK} Left panel: the distribution of CMB anisotropy spectra corresponding to stable Model 1 theories reconstructed from the H2F DE density and the relative differences w.r.t. the $\Lambda$CDM best fit model. The uncertainty due to cosmic variance around the $\Lambda$CDM best fit is shown for reference. As expected the GBD theories affect mainly the ISW effect at low $\ell$. Right panels: the linear matter power spectrum (at redshift $z=0$) for Model 1 theories reconstructed from H2F and the relative difference from the $\Lambda$CDM best fit.}
\end{figure*} 

\begin{figure*}[tbph]
\centering
\includegraphics[width = .95\textwidth]{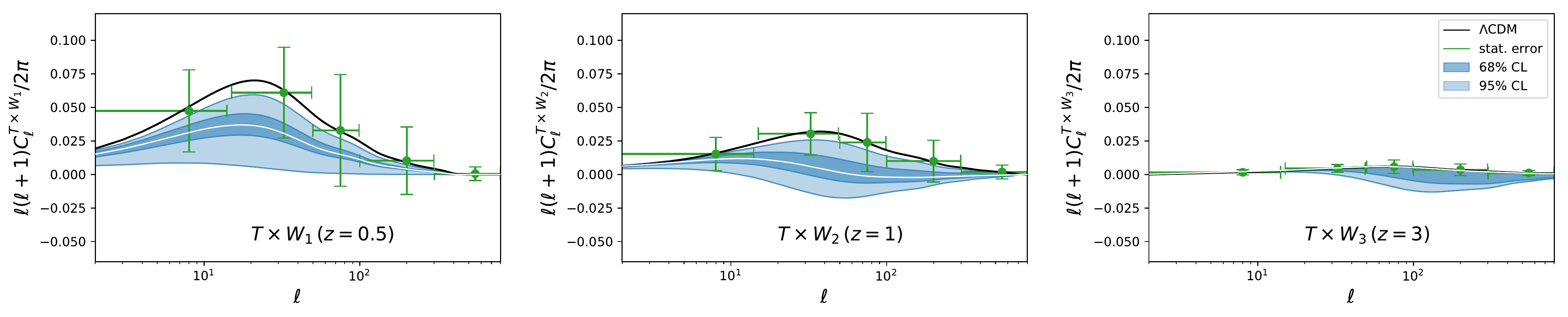} 
\caption{\label{fig:GBD_ISWGNC_CrossCorrelations} The distribution of the CMB and Galaxy Number Counts cross correlation spectra. in three redshift bins at $z=0.5$, $1$ and $3$, corresponding to the viable Model 1 theories reconstructed from H2F DE density.  The best-fit $\Lambda$CDM spectra are shown for reference. The green error bars show the uncertainty due to cosmic variance in the $\Lambda$CDM prediction in several wide bins of $\ell$. As one can see, the CMB temperature - GNC cross correlations for the GBD theories can be either positive or negative.}
\end{figure*}

\begin{figure*}[tbph]
\centering
\subfigure{
\includegraphics[width=.45\textwidth]{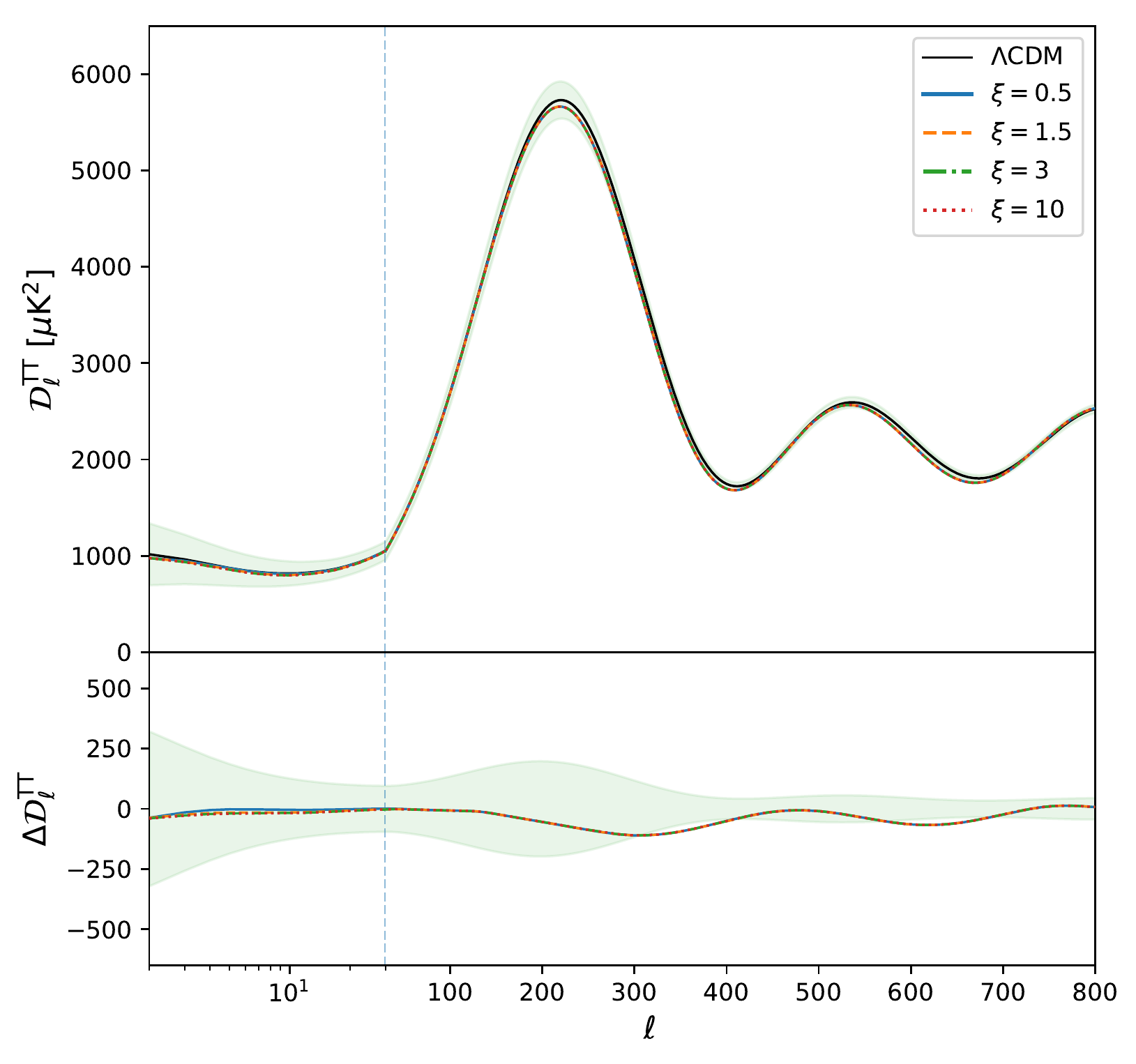}} \,
\subfigure{
\includegraphics[width=.45\textwidth]{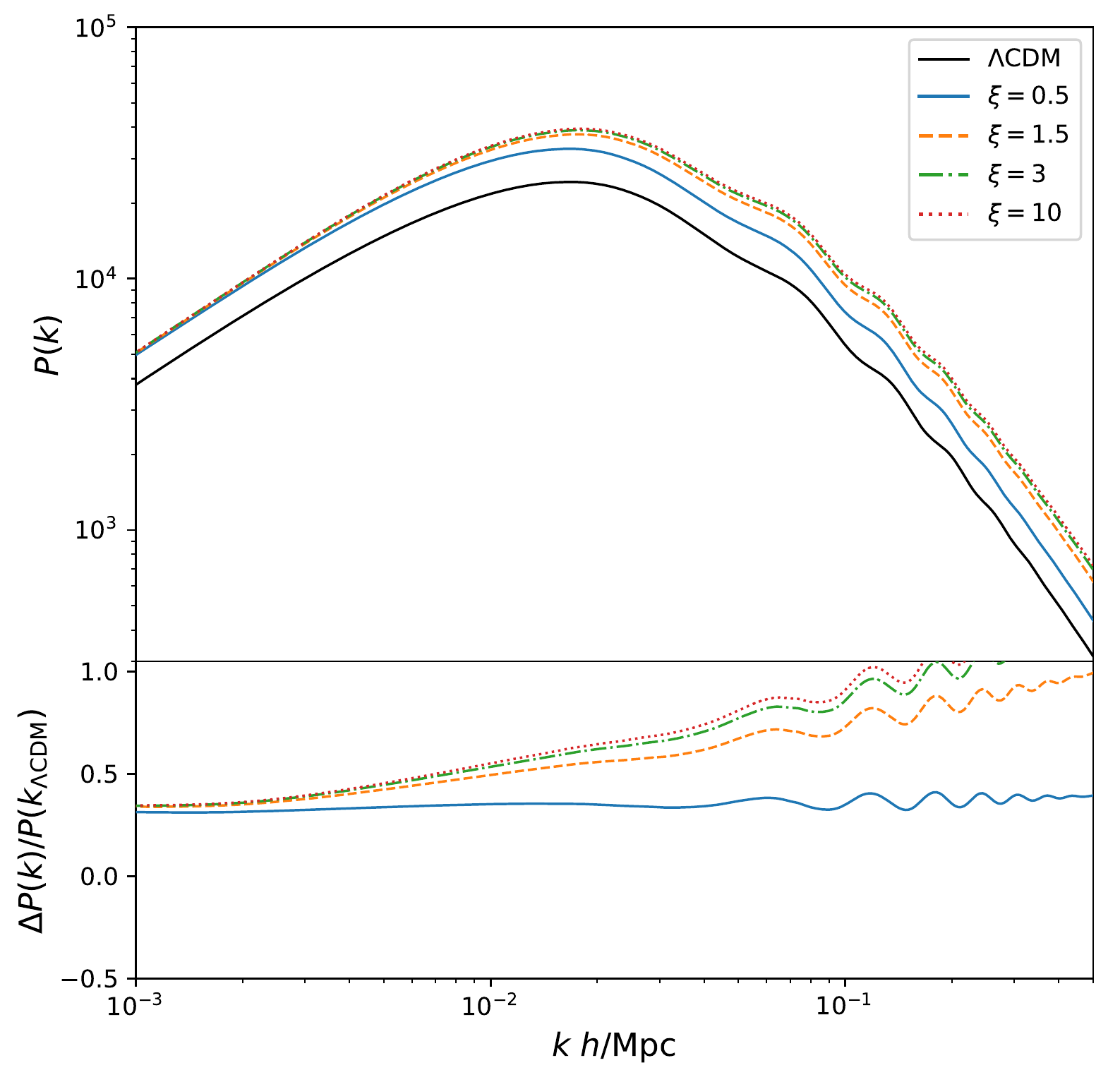}} \\
\subfigure{
\includegraphics[width=.95\textwidth]{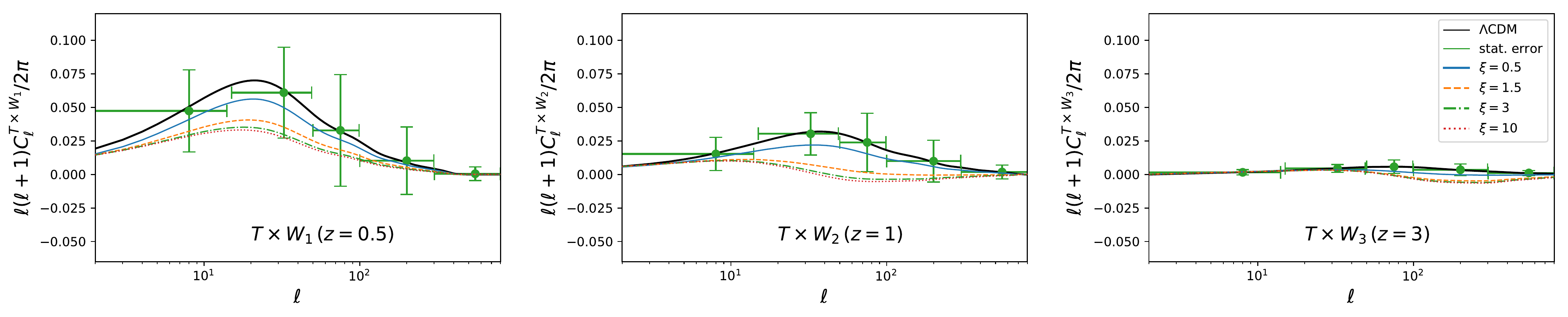}}
\caption{\label{fig:GBD_model_observables} Cosmological observables for four representative Model 1 theories reconstructed from the H2F DE density. The cosmic variance uncertainty around the $\Lambda$CDM best fit is shown for reference}
\end{figure*}

\begin{figure*}[tbph]
\centering
\subfigure{
\includegraphics[width=.45\textwidth]{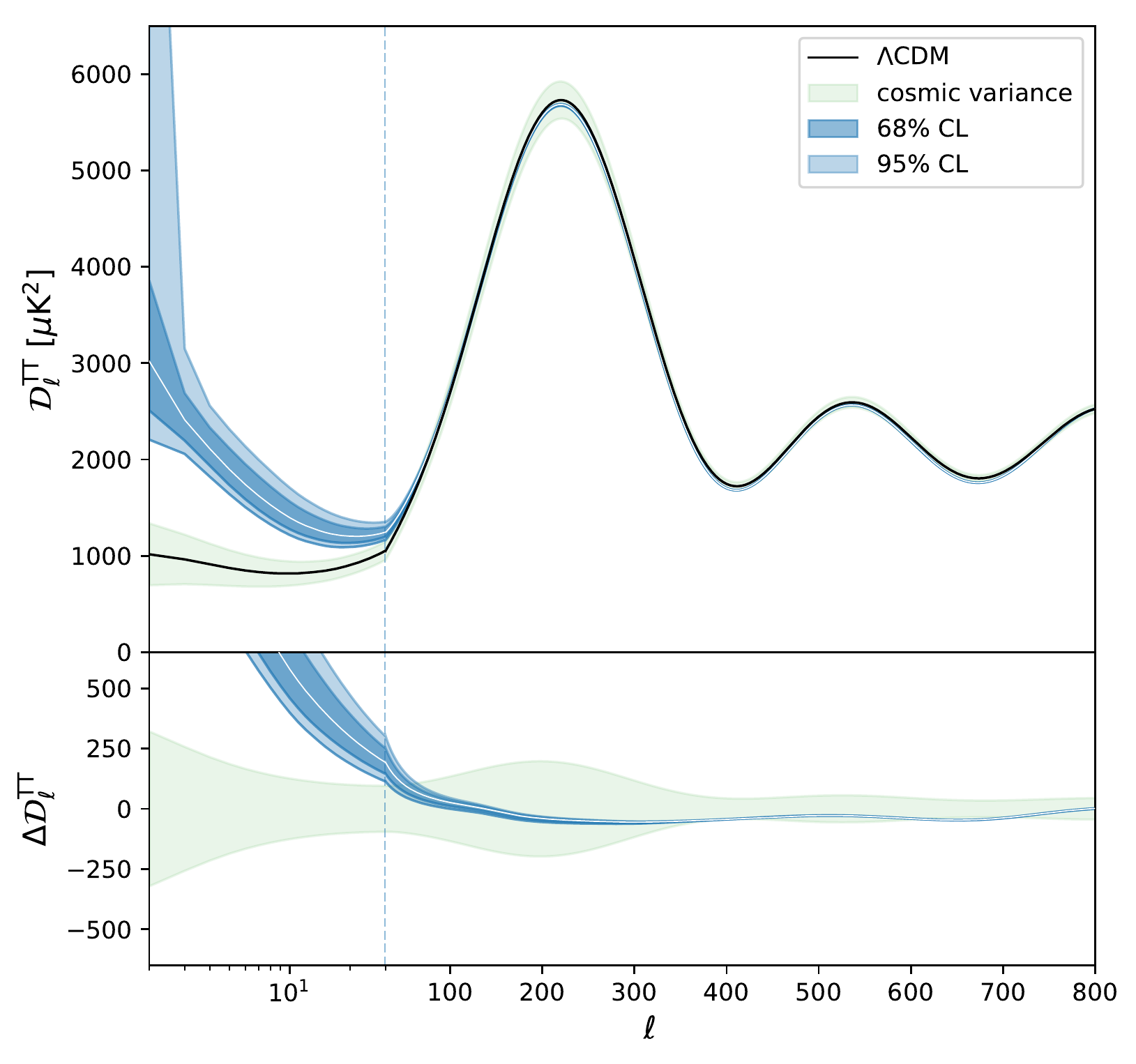}} \,
\subfigure{
\includegraphics[width=.45\textwidth]{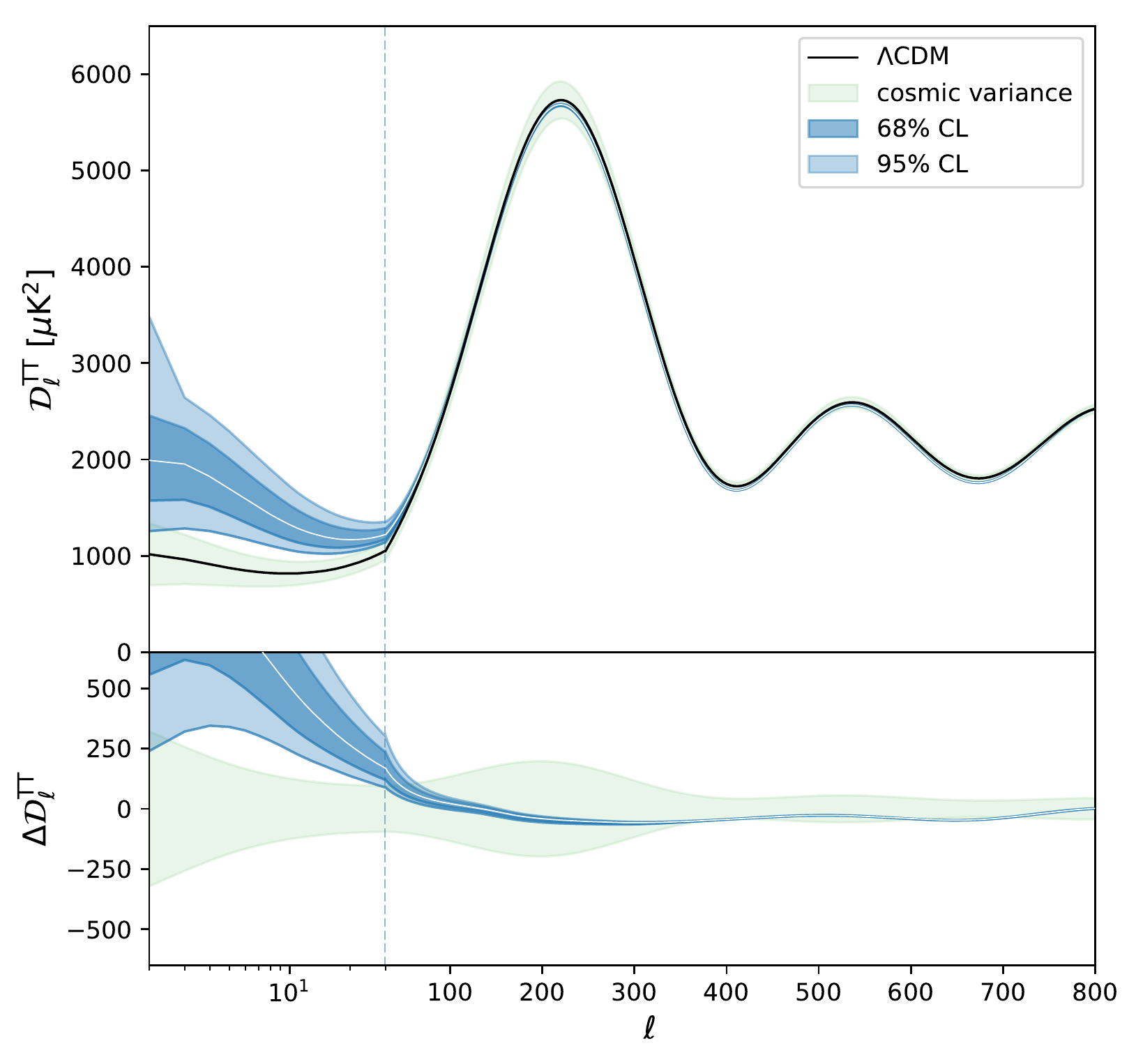}} \,
\subfigure{
\includegraphics[width=.45\textwidth]{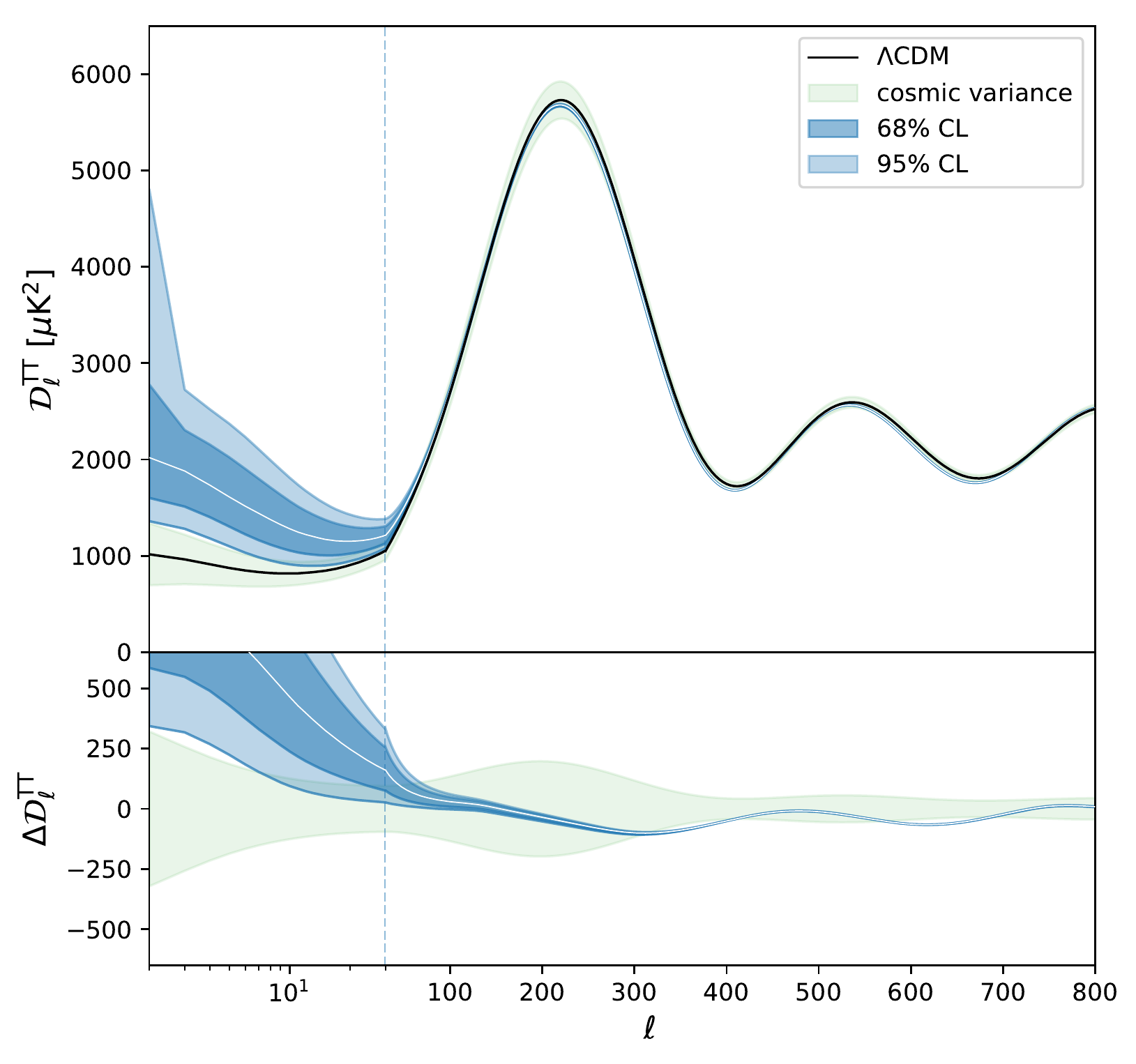}} \,
\subfigure{
\includegraphics[width=.45\textwidth]{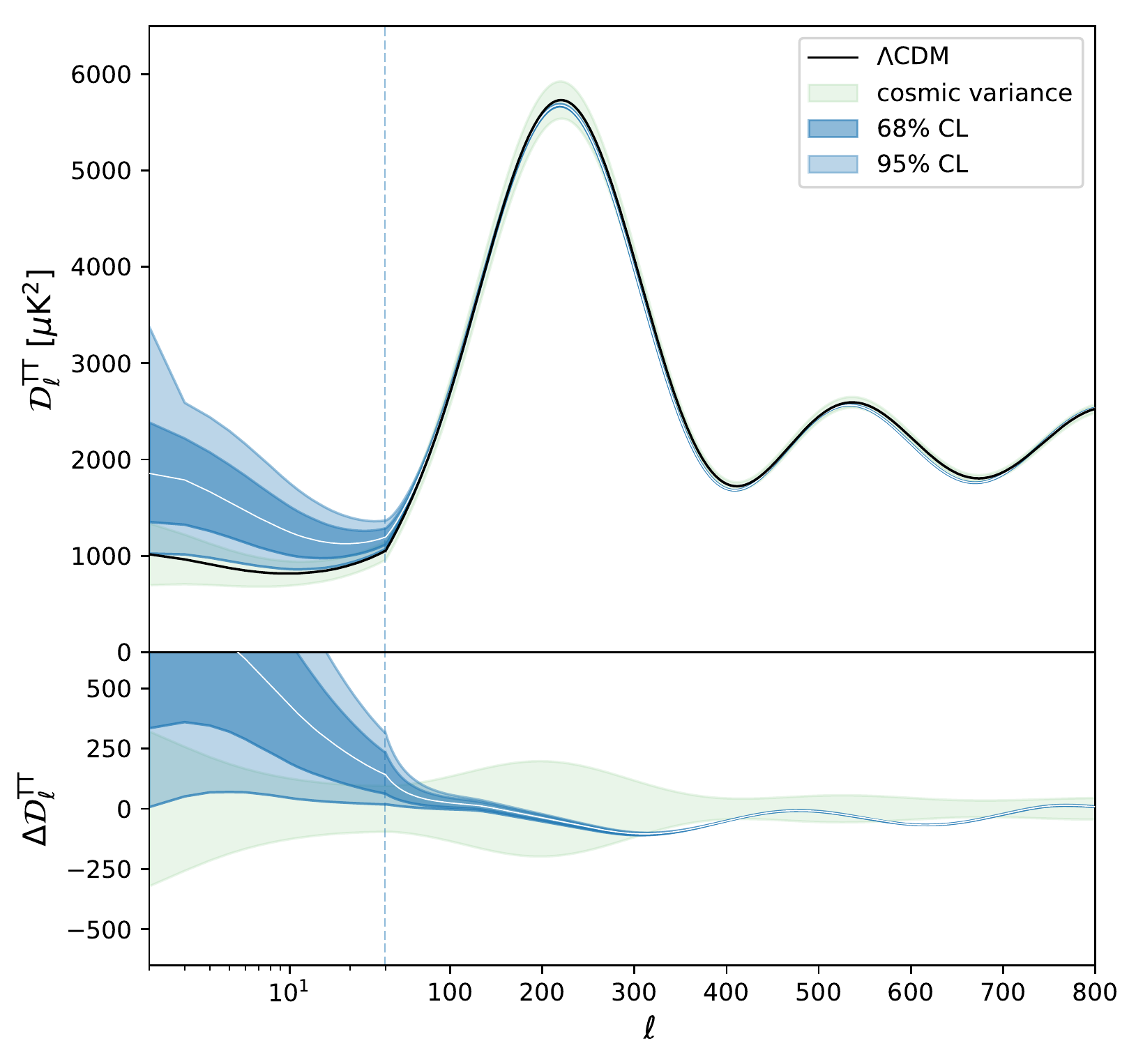}} \,
\caption{\label{fig:poly_Cls} The distribution of CMB temperature anisotropy spectra for Model 2 theories reconstructed using the H1, H1F, H2 and H2F expansion history. The best-fit $\Lambda$CDM spectra, along with the statistical error bars, are shown for reference.}
\end{figure*}

\begin{figure}[tbph]
\centering
\includegraphics[width=.45\textwidth]{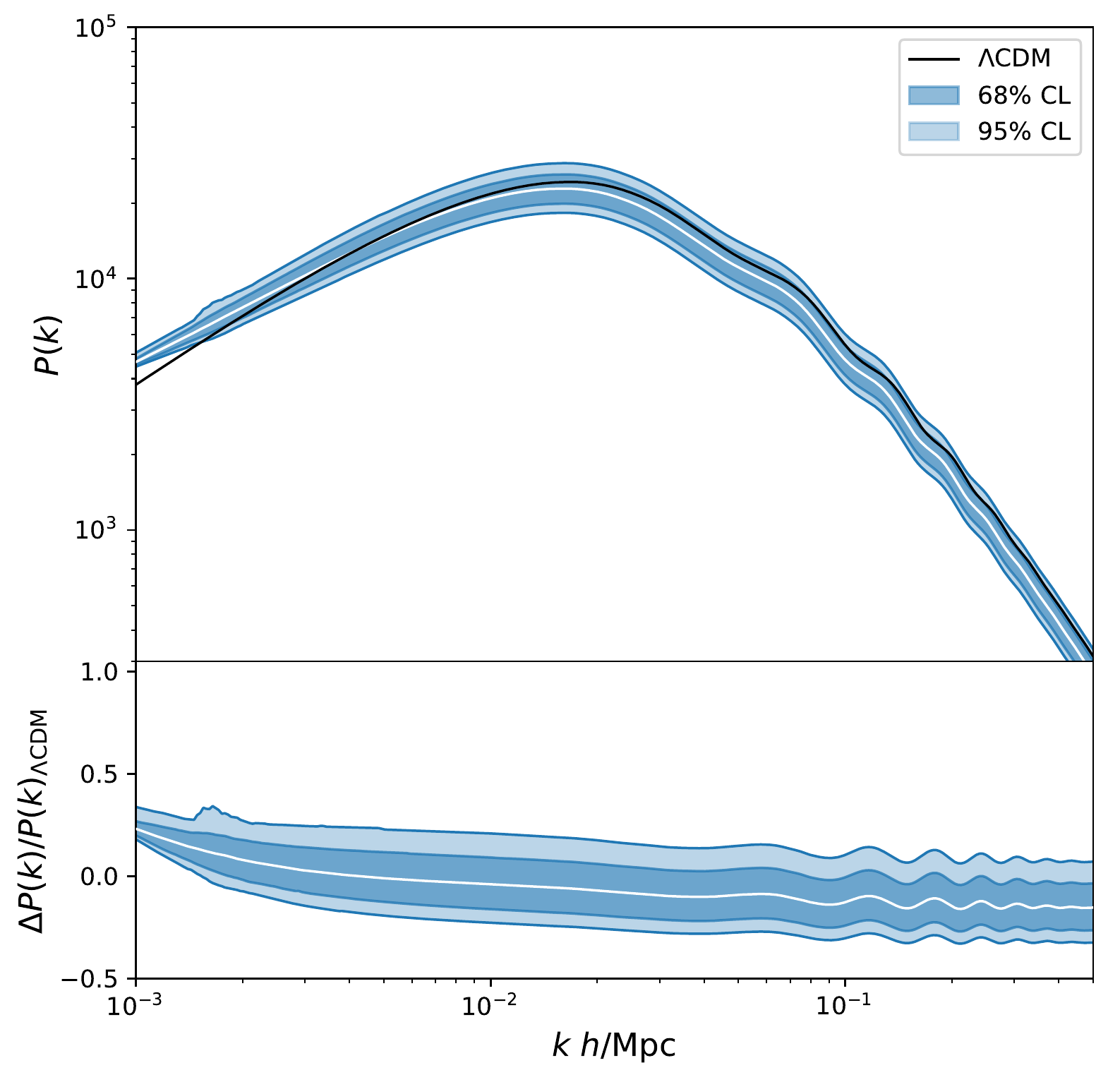}
\caption{\label{fig:poly_mpk_H2F} The matter power spectrum distribution for Model 2 theories reconstructed using the H2F expansion history. The best-fit $\Lambda$CDM spectrum is shown for reference. Distribution of spectra for H1, H1F and H2 are very similar.}
\end{figure}

\begin{figure*}[tbph]
\centering
\includegraphics[width=.95\textwidth]{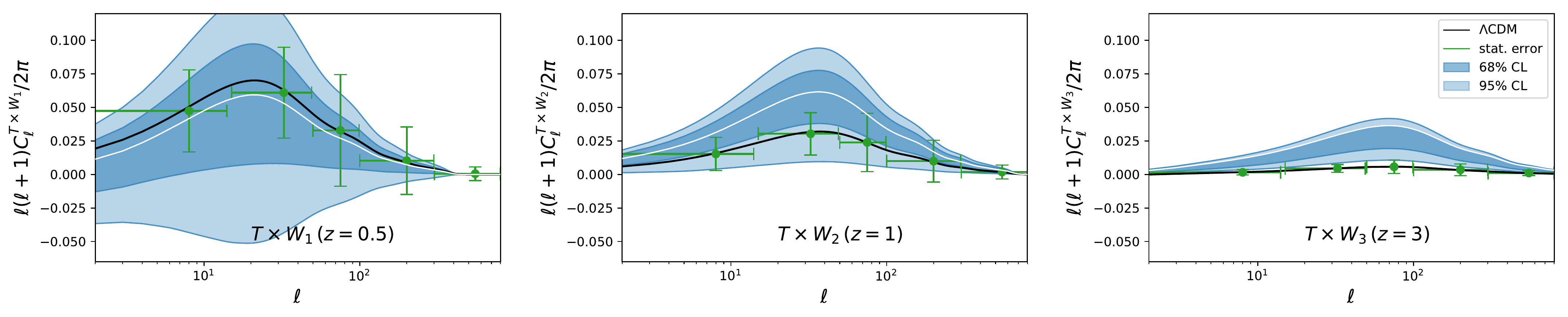} \,
\caption{\label{fig:poly_ISW} The distribution of CMB temperature and GNC cross-correlation spectra for Model 2 theories reconstructed using the H2F expansion history. The best-fit $\Lambda$CDM spectra, along with the statistical error bars, are shown for reference. The distribution of spectra for H1, H1F and H2 are qualitatively similar.}
\end{figure*}

\begin{figure*}[tbh]
\centering
\subfigure{
\includegraphics[width=.45\textwidth]{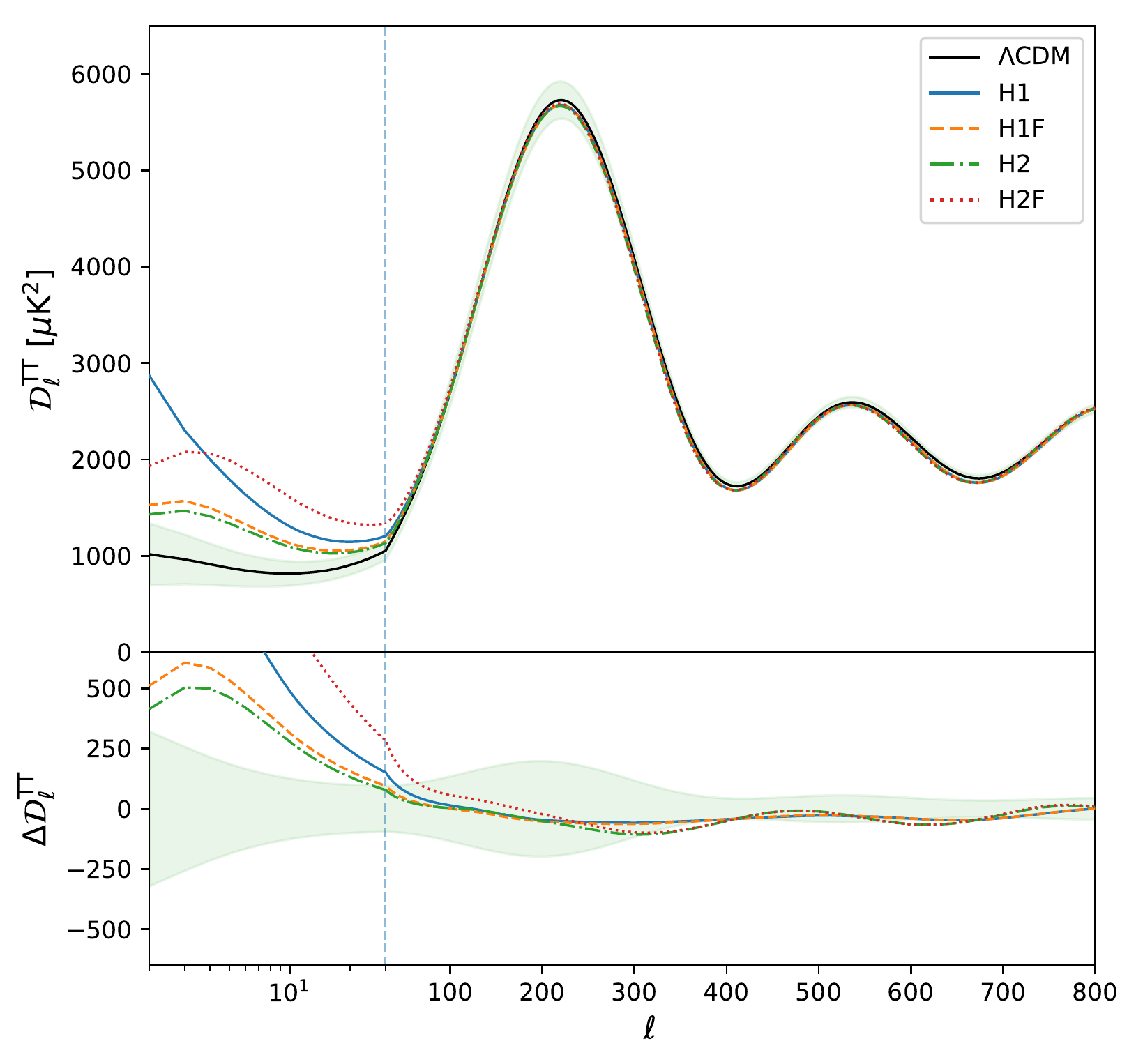}} \,
\subfigure{
\includegraphics[width=.45\textwidth]{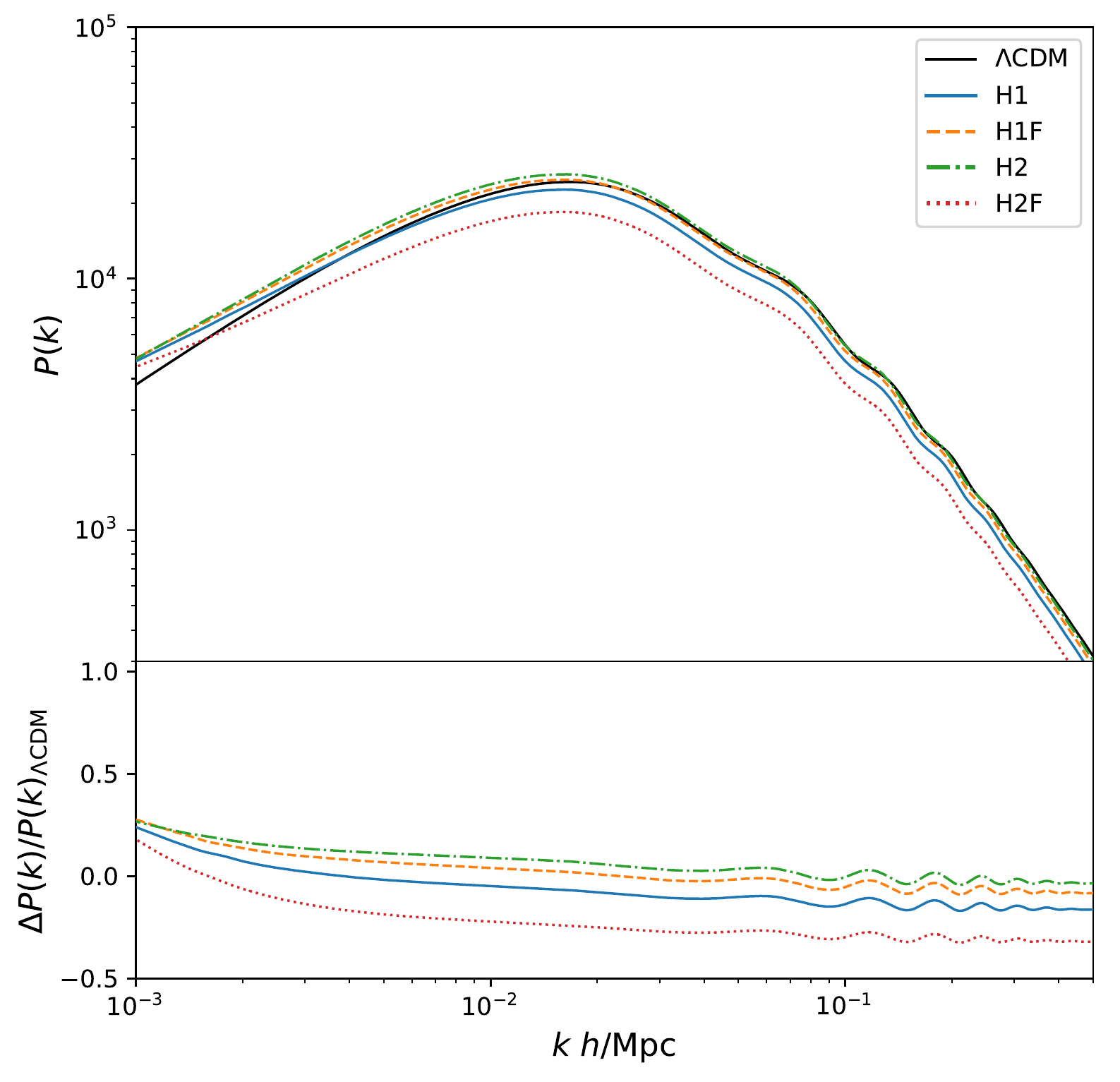}} \\
\subfigure{
\includegraphics[width=.95\textwidth]{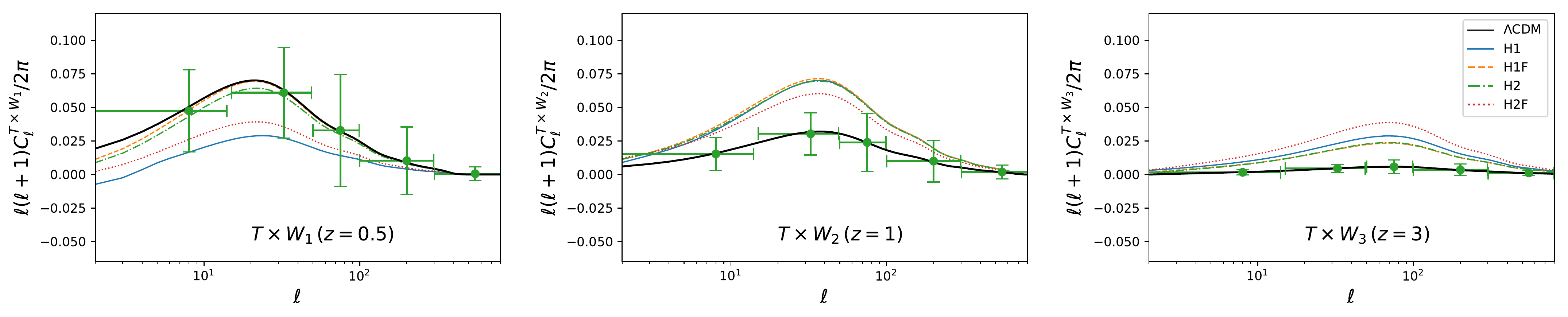}}
\caption{\label{fig:poly_cosmo_results_recon} Cosmological observables in four representative Model 2 theories reconstructed from H1, H1F, H2 and H2F expansion histories. The best-fit $\Lambda$CDM CMB spectrum, along with the statistical error bars, is shown for reference.}
\end{figure*}

Next we investigate the cosmological implications of the reconstructed GBD theories by computing the CMB anisotropy and the matter power spectra, along with the cross correlation of the CMB temperature and Galaxy Number Counts (GNC), which probes the ISW effect. It is relatively straightforward to calculate these observables using {\tt EFTCAMB} \citep{Hu:2013twa,Raveri:2014cka}\footnote{\url{https://github.com/EFTCAMB/EFTCAMB}}, which is an implementation of the effective field theory of dark energy (EFTofDE) \citep{Gubitosi:2012hu, Bloomfield:2012ff} in the popular Boltzmann solver {\tt CAMB} \citep{Lewis:1999bs}. In the EFTofDE approach, the most general action for the cosmological background and perturbations in scalar-tensor theories can be written in the unitary gauge, in which the scalar field is uniform on hypersurfaces of constant time, as an expansion in increasing rank-ordered operators invariant under spatial diffeomorphisms. The time-dependent expansion coefficients are referred to as the EFT functions. The part of the EFT action of relevance to the GBD theories is
\begin{equation}
S = \int d^4 x \sqrt{-g} \left\{ \frac{m_0^2}{2} \left[ 1 + \Omega (\tau)  \right] R + \Lambda ( \tau ) + c( \tau  ) a^2 \delta g^{00} \right\}
\end{equation}
where $\tau$ is the conformal time, $\delta g^{00} = g^{00}+1$ is the metric tensor perturbation, and $\Omega$, $\Lambda$ and $c$ are the EFT functions. The GBD theories reconstructed in the previous section can be mapped onto the EFT formalism via
\begin{align}
\Omega(a) & = F(\phi(a))-1,\\
\frac{ca^2}{m_0^2} & = \frac{1}{2} \mathcal{H}^2 (\phi^{\prime})^2, \\
\frac{\Lambda a^2}{m_0^2} & = \frac{1}{2} \mathcal{H}^2 (\phi^{\prime})^2 - Ua^2.
\label{eq:mapping}
\end{align}
With this mapping we can use EFTCAMB to compute the CMB spectra and other cosmological observables. 

As the initial time of the reconstruction $a_{\rm ini}$ is after the epoch of recombination, the only differences in the CMB anisotropy spectrum compared to the $\Lambda$CDM model can be due to the change in the expansion history, that modifies the distance to last scattering and shifts the positions of the peaks and troughs, and due to the different evolution of the gravitational potentials which affects the late-time ISW contribution to anisotropy. In the latter case, the phenomenology of GBD theories has three competing effects. First, just like in the case of $\Lambda$CDM, the accelerating expansion causes a decay of the metric potentials. In our reconstructed expansion histories the matter dominated era lasts longer, thus delaying the decay of the potentials. Secondly, the background value of the effective Newtons constant that determines the rate of gravitational clustering is $G/F(\phi)$, and can be larger or smaller than $G$, depending on the dynamics of the coupling function, correspondingly increasing or decreasing the rate at which the metric potentials evolve. Thirdly, the scalar field mediates a fifth force on scales smaller that the Compton wavelength of the field, which enhances the growth of the potentials. It is practically impossible to isolate these effects in the CMB anisotropy spectrum, since it only probes the square of the overall integral of the ISW signal. However, one can learn more by studying the correlation of CMB temperature with galaxy distribution at different redshifts \cite{1996PhRvL..76..575C,Afshordi:2004kz}. In particular, a characteristic signature of the fifth force would be a negative galaxy-CMB correlation at high redshifts, where one normally expects no ISW signal. A change in the background value of the gravitational coupling could show as either a positive or negative signal, depending on its evolution.

The CMB temperature and GNC cross correlation angular power spectrum can be written as
\begin{equation}
C_{\ell}^{Tg} = \frac{2}{\pi} \int dk \, k^2 \Delta^{\rm ISW}_{\ell} (k,\tau_0) \Delta^{\rm GNC}_{\ell}(k, \tau_0) \mathcal{P}_{\mathcal{R}}(k),
\end{equation}
where the ISW transfer function is given by 
\begin{equation}
\label{eqn:ISWsource}
\Delta^{\rm ISW}_{\ell} (k, \tau_0) = - \int_{\tau^*}^{\tau_0} d\tau \, (\dot{\Phi} + \dot{\Psi}) j_{\ell}[k(\tau_0 - \tau)],
\end{equation}
and the GNC transfer function is given by
\begin{equation}
\label{eqn:GNCTransferFunction}
\begin{split}
\Delta^{\rm GNC}_{\ell}(k, \tau_0)  = &  \int_0^{\tau_0} d \eta W(z) \frac{dz}{d \tau} b_g(\tau, k) \delta (\tau, k) j_{\ell}[k (\tau_0 - \tau)] \\
& + {\rm corrections}.
\end{split}
\end{equation}
In the above, $\Phi$ and $\Psi$ are the Newtonian gauge metric potentials in Fourier space, $\delta(k, \tau)$ is the matter density contrast, $W(z)$ is the window function that selects galaxies in the given redshift range, and $b_g$ is the galaxy bias. The term ``corrections'' in Eq.~\eqref{eqn:GNCTransferFunction} denotes collectively the redshift-space-distortion corrections, lensing terms, and other contributions suppressed by $\mathcal{H}/k$ \citep{2011PhRvD..84d3516C}. The cross-correlation spectra are then computed using the {\tt EFTCAMB} patch for {\tt CAMB sources} \footnote{The latest {\tt EFTCAMB} patch is not yet compatible with the latest {\tt CAMB}. In its last update, {\tt CAMB} and {\tt CAMB sources} have been merged, so we used the last available iteration of {\tt CAMB sources} at \url{https://github.com/cmbant/CAMB/tree/CAMB_sources}.} \citep{2011PhRvD..84d3516C, Lewis:2007kz}.

Since we are not interested in fitting the parameters of the GBD theories to data, but rather in investigating the qualitative features of the ISW effect we choose to show the cross-correlation in three Gaussian windows $W_1$, $W_2$ and $W_3$ centred at redshifts $z_1 = 0.5$, $z_2 = 1$ and $z_3 = 3$. The widths of the window functions are $\sigma_1 = 0.05, \sigma_2 = 0.1$ and $\sigma_3=0.5$. The galaxy bias $b_g$ is, in general, time and scale dependent. On large scales, relevant for the cross-correlation with CMB, one expects the scale dependence of the bias to be weak and the time dependence to have a simple polynomial dependence (see \citep{Ade:2015dva}). The bias is degenerate with the ISW amplitude, but one can calibrate it by jointly studying the GNC auto-correlations and the cross-correlations between GNC and galaxy lensing. As we are only interested in demonstrating the general features of the ISW signal, we fix the galaxy bias to $b_g=1$.

\subsection{Observables for Model 1}
\label{sec:Model1obs}

In the left panel of Fig.~\ref{fig:GBDimpact_CMB_MPK} we show the distribution of CMB temperature anisotropy spectrum ${\cal D}_{\ell} \equiv \ell (\ell + 1) (2 \pi)^{-1} C_{\ell}$ for Model 1 theories reconstructed from the H2F DE density obtained by sampling the parameter space as described in Sec.~\ref{sec:designer_gbd_fphi}. The shaded regions represent the CL regions to find ${\cal D}_{\ell}$ in the corresponding range, while the white lines show the mean values. In this sampling procedure we used the cosmological parameters obtained in the reconstruction of $X(a)$ in \cite{Wang:2018fng}, except for the parameters setting the primordial power spectrum which were not constrained in \cite{Wang:2018fng}, and for which we used the best fit $\Lambda$CDM values. The light green band shows the irreducible statistical uncertainty in ${\cal D}_{\ell}$ due to cosmic variance based on the $\Lambda$CDM model. As the ${\cal D}$ measured by Planck are cosmic variance limited over most of the cosmologically relevant $\ell$ \cite{Akrami:2018vks}, the shown uncertainty is representative of current data.

As expected, we observe a modified ISW effect at small $\ell$. The small differences in the high-$\ell$ part of the spectra are mainly due to the different distance to the last scattering surface (because of the different expansion history) which causes a shift in the peaks, and also because of the different baryon and CDM densities $\Omega_b h^2$ and $\Omega_c h^2$ in the $X(a)$ vs $\Lambda$CDM cases. These high-$\ell$ differences are well-within the cosmic variance band and would likely be accommodated by adjusting other parameters in a comprehensive MCMC parameter estimation. 
 
The right panel of Fig.~\ref{fig:GBDimpact_CMB_MPK} shows the linear matter power spectrum. First of all, one can note an overall shift upwards for the GBD theories. At early times, before DE begins to dominate the background dynamics, the Planck best-fit $\Lambda$CDM model has more DE density than the GBD models with the reconstructed DE. This means that in the GBD models the matter dominated era (MDE) lasts slightly longer than in the $\Lambda$CDM model, allowing matter to cluster more, hence the overall shift upwards of the matter power spectrum. As in the case of the CMB spectrum, we expect that this difference can accommodated by adjusting other parameters in a comprehensive fit which, however, is beyond the scope of this work. In addition to the change in the matter-DE equality, $P(k)$ is also effected by the larger $G_{\rm eff}$ and the fifth force mediated by the scalar field. This is encoded in the way the deviations from $\Lambda$CDM increase on smaller scales. Finally the oscillations that we note at $k \approx 0.1$ $h/{\rm Mpc}$ are due to the different position of the BAO scale due to a slightly different expansion history of the GBD models. 

In Fig.~\ref{fig:GBD_ISWGNC_CrossCorrelations} we show the theoretical prediction of the cross correlations for the two classes of reconstructed GBD theories from H2F. Also shown is the cosmic variance statistical uncertainty in the cross-correlation predicted by the $\Lambda$CDM model.  As one can see, in some Model 1theories, the ISW effect can become negative signalling a growing gravitational potential due to the fifth force mediated by the extra scalar field. At lower redshifts, when the effective DE becomes larger and the growth of the gravitational potential is overcome by the decay induced by the accelerated expansion, the ISW term is mainly positive. 

Some of the Model 1 theories reconstructed from the H2F expansion history are cosmologically viable, at least from the perspective of fitting the CMB spectra. Fig.~\ref{fig:GBD_model_observables} shows the cosmological observables for the four representative models whose reconstructed potentials $U(\phi)$ were shown in Fig.~\ref{fig:GBD_model_potential}. While the CMB anisotropies are almost the same for each model, they differ considerably in the clustering of matter and this is also noticeable in the cross correlations $C_{\ell}^{Tg}$ at the bottom panels. In the higher redshift windows, the larger values of the couplings constant $\xi$ drive a growth of the gravitational potentials $\Psi$ and $\Phi$ due to the fifth force mediated by the scalar field, causing a negative ISW effect. When DE eventually starts dominating the potentials stop growing and instead decay, turning the sign of the ISW effect.

\subsection{Observables for Model 2}
\label{sec:Model2obs}

Fig.~\ref{fig:poly_Cls} shows the distribution of the CMB anisotropy spectra corresponding to the sampled Model 2 theories reconstructed from the H1, H1F, H2 and H2F expansion histories. We see in all cases there is a preference for a large ISW contribution to ${\cal D}_{\ell}$. This is especially the case for H1, in which $X(a)$ is non-monotonic and has a large increase. However, since cosmic variance results in large statistical error bars at small $\ell$, there are models on the fringe of the allowed range for H1F, H2 and H2F that can be compatible with the current data. 

The Model 2 distribution of $P(k)$ is in good agreement with the data, with the best fit $\Lambda$CDM prediction being well inside the 68\% CL, as shown in Fig.~\ref{fig:poly_mpk_H2F} in the case of H2F. The matter power spectra in the cases of H1, H1F and H2F are very similar.

A very distinctive observational feature of Model 2 models is a large positive ISW signal at high redshifts, as seen in Fig.~\ref{fig:poly_ISW}. This is caused by $F > 1$ at $z \gtrsim 1$, which decreases the effective Newton's constant appearing in the Poisson equation ($G_{\rm eff} \propto G/F$) resulting in a suppression of gravitational potentials during the matter dominated epoch. The enhancement in the high redshift cross-correlation is well in excess of the cosmic variance uncertainty around the $\Lambda$CDM model and would be detectable with the next generation large scale structure surveys such as DESI, LSST and Euclid. We note that recent redshift space distortion measurements slightly favour a lower value of $G_{\rm eff}$ \cite{Zhao:2018jxv}. 

In Fig.~\ref{fig:poly_cosmo_results_recon} we show observables corresponding to the four models in Fig.~\ref{fig:reconstructed_gbd_poly} representing reconstructions using H1, H1F, H2 and H2F. We can see the similarity in general trends, with features being the most pronounces in the case of H1 and less so for H2F. However, in all cases, there is a large positive ISW signal at high redshifts which would be a smoking gun of GBD models with a non-monotonic $F(\phi)$.

\section{Conclusions}
\label{sec:summary}

Current observations favour an increasing effective DE density, corresponding to an effective DE EOS that is less than $-1$  \citep{Zhao:2017cud,Wang:2018fng}. Such apparently phantom behaviour of DE can also occur in GBD theories, as a manifestation of the additional interaction mediated by the scalar field.

We have set up a reconstruction method to design the Lagrangians of GBD type scalar-tensor theories corresponding to expansion histories extracted using the latest data probing the background \citep{Wang:2018fng}. We then examined the viability of such designer GBD theories, both in terms of their stability and their ability to predict acceptable cosmological observables.

We found that a large increase in the effective DE density, or the apparent oscillatory dynamics also favoured by the data, are difficult to accommodate within a GBD theory with a monotonically evolving coupling function, such as $F(\phi) \propto \exp({\xi\phi})$. However, allowing for an arbitrary $F(a)$, parametrized in terms of a polynomial expansion, results in GBD theories capable of fitting current CMB and matter power spectra. 

We find that, in viable models, $F(a)$ increases at high redshifts before decreasing at more recent epochs, leading to a smaller effective gravitational coupling $G_{\rm eff}$ at redshifts $z \gtrsim 1$, and a larger $G_{\rm eff}$ at $z<1$. This leads to a robust prediction of a large positive ISW signal at $z > 1$, which would be readily detectable through CMB-galaxy cross-correlation using high redshift sources from DESI, LSST and Euclid.

In our analysis, we opted to provide functions $F(\phi)$ or $F(a)$, and reconstruct the potential $U(\phi)$. One could, alternatively, opt to find $F(\phi)$ for a given $U(\phi)$. We expect that, regardless the choice, the main conclusion about the key role of the ISW effect in falsifying GBD theories will remain the same.

The method developed here is complementary to the reconstruction of the EFT functions (including the GBD subset of the EFTofDE) from cosmological observations performed in \cite{Raveri:2019mxg}. In that work, the expansion history was reconstructed in conjunction with the scalar-tensor Lagrangian, thus only producing expansion histories that are consistent with the GBD. Our approach is different -- we start with an expansion history obtained from the data in a largely model-independent way and checked if there can be GBD theories producing it. The difference is that a joint reconstruction within the GBD framework could miss expansion histories that are difficult to accommodate with smooth EFT functions, potentially missing a hint for dynamics that would correspond to a rare realization of GBD.

Our results show that one could rule out scalar-tensor theories as the explanation of departures from the $\Lambda$CDM background expansion history using purely cosmological data sets. This is particularly important for testing theories in which the scalar field couples only to dark matter, to which the tight laboratory and solar system tests of gravity do not apply.

\acknowledgments

We acknowledge G. Papadomanolakis and S. Peirone for useful discussions and feedback on our findings. A.Z. and L.P. are supported by the Natural Sciences and Engineering Research Council (NSERC) of Canada. A.S. acknowledges support from the NWO and the Dutch Ministry of Education, Culture and Science (OCW), and from the D-ITP consortium, a program of the NWO that is funded by the OCW. GBZ is supported by the National Key Basic Research and Development Program of China (No. 2018YFA0404503), and by NSFC Grants 11720101004 and 11673025. Y.W. is supported by the Nebula Talents Program of NAOC. This research was enabled in part by support provided by WestGrid and Compute Canada.


%

\end{document}